\documentclass[10pt,journal,compsoc]{IEEEtran}

\ifCLASSOPTIONcompsoc
  \usepackage[nocompress]{cite}
\else
  \usepackage{cite}
\fi
\usepackage{cutwin}
\ifCLASSINFOpdf
\else
\fi
\usepackage{multirow}
\usepackage{subfigure}
\usepackage{subcaption}
\usepackage{pifont}
\usepackage{epsfig}
\usepackage[ruled,vlined]{algorithm2e}
\usepackage{tikz}
\usepackage[none]{hyphenat}
\usepackage[dvipsnames]{xcolor}
\usepackage{colortbl}
\usepackage{balance} 

\usepackage[numbers,sort&compress]{natbib}
\newcommand\MYhyperrefoptions{bookmarks=true,bookmarksnumbered=true,
pdfpagemode={UseOutlines},plainpages=false,pdfpagelabels=true,
colorlinks=true,linkcolor={black},citecolor={black},urlcolor={black},
pdftitle={Bare Demo of IEEEtran.cls for Biometrics Council Journals},
pdfsubject={Typesetting},
pdfauthor={Michael D. Shell},
pdfkeywords={Biometrics Council, IEEEtran, journal, LaTeX, paper,template}}
\ifCLASSINFOpdf
\usepackage[\MYhyperrefoptions,pdftex]{hyperref}
\else
\usepackage[\MYhyperrefoptions,breaklinks=true,dvips]{hyperref}
\usepackage{breakurl}
\fi
\begin{document}

\title{Detecting LLM-Assisted Academic Dishonesty using Keystroke Dynamics}
\author{ Atharva Mehta*, Rajesh Kumar*, Aman Singla, Kartik Bisht, Yaman Kumar Singla, Rajiv Ratn Shah
\IEEEcompsocitemizethanks{\IEEEcompsocthanksitem *equal contributions. Atharva Mehta (atharva.mehta@mbzuai.ac.ae) is with MBZUAI, Rajesh Kumar (rajesh.kumar@bucknell.edu) is with Bucknell University, Yaman Kumar Singla (ykumar@adobe.com) is with Adobe Media and Data Science Research, Aman Singla (aman.singla@midas.center), Kartik Bisht (kartik.bisht@midas.center), and Rajiv Ratn Shah (rajivratn@iiitd.ac.in) are with IIIT Delhi, India. }
}

\IEEEtitleabstractindextext{%
\begin{abstract}
The rapid adoption of generative AI tools has heightened concerns regarding academic integrity, as students increasingly engage in dishonest practices by copying or paraphrasing AI-generated content. Existing plagiarism detection systems, which rely primarily on text-intrinsic features, are ineffective at identifying AI-assisted or paraphrased submissions. Our prior conference work \cite{kundu2024keystroke} introduced a behavioral detection approach that leverages how text is produced, captured through keystroke dynamics, in addition to what is written, enabling discrimination between genuine and assisted writing. That study, conducted on keystroke data from 40 participants, demonstrated promising performance. This paper substantially extends and systemizes the prior work by: (1) expanding the dataset with 90 additional participants and introducing an explicit paraphrasing condition to model realistic plagiarism strategies; (2) formalizing a threat model and evaluating detection under adversarial and deception-oriented scenarios; and (3) performing a comprehensive empirical comparison against state-of-the-art text-only detectors and human evaluators. Experimental results demonstrate that keystroke-based models significantly outperform text-based approaches in practical deployment settings, while revealing limitations under more challenging adversarial conditions.

\end{abstract}

\begin{IEEEkeywords}
Keystroke dynamics, academic dishonesty detection, LLM-assisted writing \\ 

©2022 IEEE. Personal use of this material is permitted. Permission from IEEE must be obtained for all other uses in any current or future media, including reprinting/republishing this material for advertising or promotional purposes, creating new collective works, for resale or redistribution to servers or lists, or reuse of any copyrighted component of this work in other works. To appear in \\ 
\underline{IEEE Transactions on Biometrics, Behavior, and Identity Science, 2026.}
\end{IEEEkeywords}}
\maketitle

\IEEEdisplaynontitleabstractindextext


%
\IEEEpeerreviewmaketitle
\section{Introduction}
\label{sec:introduction}
With the advent of large language models (LLMs) such as ChatGPT \cite{ChatGPT}, Gemini \cite{geminiteam2025}, Claude \cite{Claude}, and LLaMA \cite{grattafiori2024llama}, alongside the proliferation of advanced copywriting and copyediting tools, it has become increasingly challenging for academics and educational testing agencies to determine whether a text is authored by a human or generated by AI \cite{ChatGPTOnEducation}. This difficulty has exacerbated academic dishonesty, including plagiarism (presenting another's work as one's own), cheating (using unauthorized tools during exams), and collusion (unauthorized collaboration). Studies show that human accuracy in distinguishing human-written text from GPT-generated text is only marginally better than random guessing (65\%) \cite{clark2021all,ippolito2019automatic, roh2025llm}. This level of accuracy falls significantly below acceptable thresholds for high-stakes educational assessments \cite{dugan2023real}. Moreover, while humans often associate poor organization and grammatical errors with human authorship, prompt engineering can easily produce AI-generated text that mimics these traits \cite{yan2023detectionEssayWriting,clark2021all,ippolito2019automatic, jemma2025how}. Interestingly, human raters frequently struggle to articulate the rationale behind their judgments. Explanations are often limited to vague statements, such as the text \textit{rambles in a way that makes sense} or is \textit{too natural to be AI-generated} \cite{clark2021all,ippolito2019automatic}. This highlights the inherent difficulty of reliably detecting AI-generated content using subjective human evaluation alone. 


Recognizing the critical challenge of detecting AI-generated text, recent research has focused on developing automated detection methods. These approaches predominantly rely on stylistic features of text, analyzed using models such as n-grams, handcrafted features, and transformers \cite{fagni2021tweepfake,solaiman2019release,uchendu2020authorship,zellers2019defending,frohling2021feature}, or perplexity-based metrics \cite{gehrmann2019gltr,kumar2023automatic,singla2022minimal,lavergne2008detecting}. Additionally, several commercial tools, including Zerogpt \cite{ZeroGPT}, Quillbot \cite{quill}, GPTZero \cite{gptzero}, and Writefull \cite{Writefull}, claim high accuracy in distinguishing human-written text from that generated by large language models (LLMs). However, it is noteworthy that major educational boards have not yet officially tested or approved these tools. Despite the lack of formal endorsement, individual classes and schools have started experimenting with these tools on their students. This has led to numerous reports of disputes involving students and parents challenging allegations of cheating based on false positives \cite{hindu2024llm,coffey2024professors,merod2023turnitin,young2024grammarly,coley2023guidance,tenbarge2024ai}. 

In this paper, we propose an alternative approach in which, in addition to submitting a text for grading in an assignment or exam, a human also submits their behavioral patterns (e.g., keystroke logs) as proof of authorship. The key idea is that, while producing and editing any work, humans generate numerous behavioral signals, including keystroke logs, gaze patterns, brain activation patterns, and electrodermal signals. These signals can serve as evidence of the human writing process and effort, thereby helping us distinguish between LLM-generated and human-generated text. While other behavioral signals like those from the brain, skin, and eyes require special devices, which are often privacy-invasive, keystroke logs are much easier to collect (require just a client-side JavaScript installation) and do not require any special equipment beyond a simple keyboard or keypad \cite{plurilockKeystrokeDynamicsDiffMimic}. Furthermore, while text style can be mimicked, studies on keystrokes show that it is difficult to emulate someone’s typing patterns in general \cite{plurilockKeystrokeDynamicsDiffMimic, khan2018augmented, MimicryAttacksKeystrokes}. That means it would be difficult for people to hire someone (aka contract cheating) to commit cheating if we adopt keystroke-based detection of cheating or plagiarism. It has been found that by pressing only $100$ keys, one can authenticate/identify a person from a database of $1000$ users~\cite{acien2021typenet}. In this work, we investigate whether it is possible to integrate behavioral signals captured through keystroke logs with text stylistic features to detect whether a human made an effort to produce a piece of text. We demonstrate that a model integrating text and keystroke logs can achieve $F_1$ scores between 64\% and 96\% on unseen writers and unseen text (for which the model has received no signals from those humans or the same text during training). 

By integrating behavioral signals into our model, we leverage the process of text production \cite{hayes1981uncovering,matsuhashi1981pausing,van2008pause} rather than relying solely on the final product (the text itself). This approach provides a more nuanced understanding of writing behaviors. While straightforward cases, such as directly copying and pasting LLM-generated text, can be effectively mitigated by disabling such functionalities in the submission interface, our findings demonstrate that models incorporating keystroke logs achieve high accuracy in detecting both \textit{transcribed} text (verbatim typing from a source) and \textit{paraphrased} text (typing with modifications to the source content). 

Our conference  \cite{kundu2024keystroke} paper: (a) Proposed a new keystroke dataset collected from $40$ students completing both \textit{bona fide} (independent writing) and \textit{AI-assisted} (using ChatGPT) tasks, covering opinion- and fact-based questions, and (b) adapted and evaluated TypeNet \cite{acien2021typenet} architecture for plagiarism detection on the proposed and two publicly available (repurposed) datasets \cite{banerjee2014_emnlp, 7823894}, across user, keyboard, context (topic), and dataset-specific/agnostic scenarios, showing the promise of keystroke signals in distinguishing AI-assisted and genuine writing. 
 
This paper systematically expands the conference version in the following dimensions:

\begin{itemize}
    \item It extends the dataset by adding $90$ new participants and introducing a \textit{paraphrasing} mode to better reflect realistic plagiarism strategies.
    \item It proposes a formal threat model and a mechanism for generating adversarial keystroke sequences, and uses them to evaluate model robustness under deception scenarios.
    \item It conducts a comprehensive empirical comparison against text-only detectors and human evaluators.
\end{itemize}  


The remainder of the paper is structured as follows: Section \ref{sec:related_work} reviews related work, Section \ref{sec:materials_and_methods} outlines the materials and methods, Section \ref{sec:results_discussions} presents the results and discussions, and Section \ref{sec:conclusion_future_work} concludes the paper with a summary and future research directions. 


\section{Related Work}
\label{sec:related_work}

Detecting if a text is written by a human or generated automatically is a longstanding problem with applications in academic dishonesty detection \cite{foltynek2019academic}, fake news identification \cite{mridha2021comprehensive}, fake product review identification \cite{ott2011finding}, and spamming and phishing \cite{crawford2015survey}. These problems have been worsened by LLMs. While recognizing this problem, OpenAI released a report on their initial generative model, GPT-2, and presented manual and automated experiments to detect GPT-generated text using machine-learning and text-based transformer models~\cite{solaiman2019release}. Furthermore, it has been shown that humans can identify GPT-generated text \cite{ippolito2019automatic,clark2021all}, fake news articles, fake product reviews \cite{adelani2020generating}, and fake comments \cite{weiss2019deepfake} generated automatically only at a chance level. Interestingly, this finding aligns with other domains of behavioral research, where it has been demonstrated that humans perform just above chance in predicting various types of behavior \cite{tan2014effect,isola2013makes,si2023long,singh2024measuring,tetlock2017expert}.

Given the wide applicability of machine-generated text detection, two broad approaches have been developed to determine whether a text is machine-generated. The first approach is to build classical machine learning-based classification models that take text as input and classify it as machine-generated based on features such as entropy, perplexity, n-gram frequency, \textit{etc.} \cite{lavergne2008detecting,gehrmann2019gltr,mitchell2023detectgpt}. The second approach is to change the text generation model to induce specific patterns in the generated text. This is referred to as watermarking \cite{atallah2001natural,kirchenbauer2023watermark}. 

While the prior literature has reported the success of these classification techniques, they are unsuccessful in practice, primarily due to their likelihood of detecting a writing sample as plagiarized when it is not \cite{hindu2024llm,coffey2024professors,merod2023turnitin,young2024grammarly,coley2023guidance,tenbarge2024ai}. Advancements in writing tools and generative AI models have made it possible to generate a diversity of writing styles. Additionally, most previous literature focuses on detecting fully AI-generated text. This overlooks several crucial copyediting workflows, including automated generation, followed by manual or automated style enhancement, and rewriting. These workflows are important parts of the writing process \cite{hayes1981uncovering}. 

Therefore, in our conference paper \cite{kundu2024keystroke}, we proposed moving beyond text-intrinsic detection methods and accounting for the text-production process. In particular, we used keystroke dynamics, an established method of capturing the text-production process. The keystrokes capture several behavioral signals while ideating, planning, drafting, writing, revising, and editing. Literature \cite{trezise2019contract_cheating_detecting, agarwal-nancy-contract} on keystroke dynamics suggests that it provides a complementary view of a user's style beyond linguistic signals. They help us identify the cognitive processes at play during the writing process that distinguish it from cases in which a different cognitive process is employed, for example, when transcribing or paraphrasing a text written by someone else or GenAI. Early research in psychology and educational science had indicated that keystroke logs could help infer cognitive signals during writing processes \cite{van2009keystroke,wengelin2006examining,baaijen2012keystroke,ekman2003darwin,vizer2009automated}. Prior literature has also shown that cognitive load captured via behavioral signals can improve AI model performance \cite{plank2016keystroke,singh2024llava,khurana2023synthesizing,sood2020improving}.
 
This paper thus primarily extends and systemizes the ideas presented in our conference paper \cite{kundu2024keystroke}. In particular, it extends the dataset, typing modes, threat modeling, classification methods, and evaluation under deception scenarios. 

\section{Materials and methods}
\label{sec:materials_and_methods}

\subsection{Datasets}
We use four keystroke datasets in this study. Two of them were created to study GenAI-assisted plagiarism and are referred to as IIITD-BU (Transcribed) and IIITD-BU (Paraphrased). The Transcribed dataset captures direct copying and typing of large language model (LLM) outputs, whereas the Paraphrased dataset captures reworded LLM responses. The remaining two, viz. Stony Brook University (SBU) and SUNY Buffalo (Buffalo) datasets were repurposed to suit the problem context \cite{banerjee2014_emnlp,7823894}. These two datasets provide explicit annotations for fixed and free typing modes, which map directly to assisted (plagiarized) and bona fide (non-plagiarized) writing scenarios. Table \ref{tab:comparison} summarizes the key characteristics of all four datasets along with the descriptions provided below:

\subsubsection{SBU Dataset}
The SBU dataset \cite{banerjee2014_emnlp} was created using Amazon Mechanical Turk and contains texts on restaurant reviews, gun control, and gay marriage. Each participant produced both truthful texts reflecting their genuine opinions and deceptive texts contradicting those opinions, with all texts exceeding $100$ words. Data were collected under two writing modes: \textit{free writing}, in which participants composed original responses to prompts, and \textit{fixed writing}, in which participants retyped texts they had previously authored.

Keystroke logging captured \textit{KeyDown} and \textit{KeyUp} events, and high-resolution timestamps. Both printable keystrokes (e.g., alphanumeric characters) and non-printable keys (e.g., Backspace and Delete) were recorded, enabling analysis of pauses, revisions, and editing behavior. Copy-and-paste functionality was disabled, and participants completed tasks in two counterbalanced sequences: truthful followed by deceptive writing, and the reverse order.

The distinction between free and fixed writing provides a controlled contrast between original text generation and transcription-dominated typing behavior. Free writing captures variability associated with text composition, including pauses, revisions, and temporal irregularities, whereas fixed writing reflects behavior driven primarily by recall or transcription. Differences in temporal dynamics, revision frequency, and typing regularity provide discriminative behavioral features for such analysis. These properties make the SBU dataset suitable for research on academic dishonesty detection. 

\begin{table}[htp]
\centering
\caption{Comparison of keystroke datasets used in this study. SBU and Buffalo are publicly available datasets, while IIITD-BU (Transcribed) and IIITD-BU (Paraphrased) were collected for this work (includes conference paper \cite{kundu2024keystroke}). \textit{Context} indicates variation in writing topics, \textit{Keyboard} denotes the presence of multiple keyboard types, \textit{Cognitive} indicates tasks with varying cognitive demands, and \textit{AI-generated} denotes tasks involving AI-generated content.}

\vspace{0.15in}
\begin{tabular}{|l|c|c|c|c|}
\hline
\textbf{Features} & \textbf{SBU} & \textbf{Buffalo} & \shortstack{\textbf{IIITD-BU} \\ \textbf{Transcribed}} & \shortstack{\textbf{IIITD-BU} \\ \textbf{Paraphrased}} \\
\hline
Users & 196 & 157 & 34 & 90\\ 
\hline
Free/bonafide & \ding{51} & \ding{51} & \ding{51} & \ding{51}\\ 
\hline
Fixed/assisted & \ding{51} & \ding{51} & \ding{51} & \ding{51}\\ 
\hline
Context & \ding{51} & \ding{55} & \ding{51} & \ding{51}\\ 
\hline
Keyboard & \ding{55} & \ding{51} & \ding{55} & \ding{55}\\ 
\hline
Cognitive & \ding{55} & \ding{55} & \ding{51} & \ding{51}\\ 
\hline
AI-generated & \ding{55} & \ding{55} & \ding{51} & \ding{51}\\ 
\hline
\end{tabular}
\label{tab:comparison}
\end{table}

\subsubsection{Buffalo Dataset}
The Buffalo keystroke dataset \cite{7823894} was collected from $157$ participants over three sessions, with an average inter-session interval of $28$ days, and was originally designed for continuous authentication research. Data collection involved two task categories: \textit{fixed text transcription}, in which participants retyped excerpts from Steve Jobs' Stanford commencement speech, and \textit{free-style typing}, which included spontaneous and semi-structured writing activities such as expressing personal opinions, describing visual scenes, and performing routine tasks including email composition and web interaction. The contrast between fixed- and free-type tasks make this dataset suitable for this study. 

Data were collected using both consistent baseline keyboards and a keyboard-variability subset to capture device-induced variation. Keystroke logging recorded \textit{KeyDown} and \textit{KeyUp} events, and mouse coordinates, enabling analysis of temporal dynamics, editing behavior, and mouse interactions. 

\subsubsection{IIITD-BU (Transcribed) Dataset}
This dataset was collected to complement existing keystroke dynamics datasets by introducing controlled variation in typing behavior induced by Generative AI assistance (ChatGPT). Data collection consisted of two sessions. In \textit{Session~1}, participants typed responses without any AI assistance to capture bona fide typing behavior. In \textit{Session~2}, participants transcribed content generated by ChatGPT, simulating plagiarism through direct copying. In both sessions, participants answered opinion-based and fact-based questions, with a minimum requirement of $300$ typed characters per question. The protocol targeted two levels of cognitive load inspired by Bloom’s revised taxonomy \cite{anderson2001taxonomy}: fact-based questions corresponding to lower cognitive levels (remembering, understanding, and applying), and opinion-based questions corresponding to higher cognitive levels (analyzing, evaluating, and creating). The dataset records \textit{KeyDown} and \textit{KeyUp} events along with precise timestamps. Data were collected in a classroom setting, and participants used their own laptops, resulting in variability in physical keyboard layouts.


\subsubsection{IIITD-BU (Paraphrased) Dataset}
IIITD-BU (Paraphrased) is essentially and extension of the IIITD-BU (Transcribed) targeting intelligent plagiarism scenarios in which participants incorporate AI-generated content but paraphrase it. Literature \cite{hayes1981uncovering,matsuhashi1981pausing,van2008pause} suggests that paraphrasing exhibits distinct behavioral patterns. We hypothesized that paraphrasing behavior would reflect differently into keystroke patterns than transcription. 

The participants in IIITD-BU (Paraphrased) Dataset were primarily students enrolled in a course on large language models at IIIT Delhi. To maintain consistency with the IIITD-BU (Transcribed) dataset, one question targeted lower cognitive levels (recall and understanding), while the other required higher-order processes such as analysis and evaluation. The questions asked include (a) What is a large language model (LLM), and how does it work? Explain how an LLM such as ChatGPT is trained, how transformer architectures are used, and provide an example application (Minimum $250$ words), and (b) Analyze the strengths and weaknesses of using large language models, such as ChatGPT, to answer homework questions. Propose one improvement that could increase their usefulness in educational settings and justify your choice (Minimum $500$ words). 

Participants completed two sessions. In the first session, they answered the questions independently, relying solely on their own knowledge and reasoning, establishing a baseline for bona fide typing behavior. In the second session, participants used LLMs such as ChatGPT for assistance but were explicitly instructed to paraphrase the generated content before typing.

To ensure originality, copy-and-paste functionality was disabled, and participants were required to disable auto-correct, grammar tools, and browser extensions. Both sessions recorded \textit{KeyDown} and \textit{KeyUp} events, and timestamps, enabling detailed analysis of paraphrasing behavior in comparison to transcription and independent writing.

\subsubsection{Deception Datasets} \label{sec:attack_dataset}
To evaluate the robustness of the proposed models for distinguishing assisted and bona fide typing behavior (Section~\ref{sec:design_detector}), we construct deception datasets that simulate deceptive typing while preserving the statistical characteristics of natural keystroke dynamics. The construction process follows a systematic threat modeling procedure explained in Section \label{sec:deception_threat_model}.

Essentially, we construct timing dictionaries for each user and each key or key combination. Observed KHTs and KITs are summarized using empirical means ($\mu$) and standard deviations ($\sigma$), which parameterize Gaussian distributions $N(\mu,\sigma^{2})$ from which synthetic timings are sampled.

To ensure coverage and robustness, a three-stage lookup strategy is employed during synthesis: (i) user-specific statistics, (ii) pooled statistics across users for the same key or key combination, and (iii) a global fallback if neither is available. To approximate natural editing behavior, backspace events are stochastically injected after every $\geq 6$ characters with a probability of 80\%.

Based on this process, we construct two deception threat models: (1)~\textit{user-specific attack} (At-U) and (2)~\textit{pooled attack} (At-P). The At-U model follows the full three-stage lookup procedure, whereas the At-P model excludes user-level statistics and relies solely on pooled distributions to simulate general bona fide typing behavior.  

\subsection{Design of the detectors} \label{sec:design_detector}
\begin{figure}[htp]
    \centering
    \includegraphics[width=3.2in, height= 1.66in]{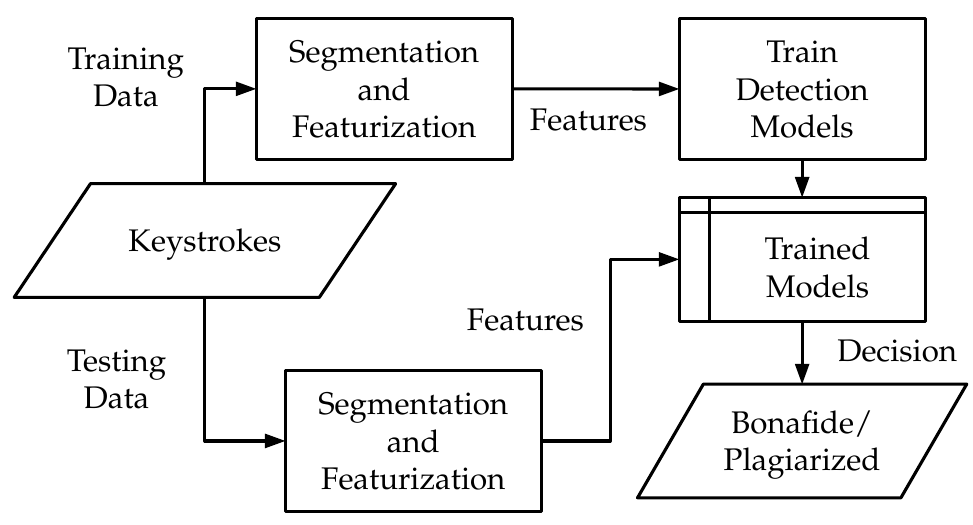}
    \caption{The plagiarism detection framework shows the input keystrokes, their segmentation, featurization, training of the models, and their evaluations on the test keystrokes for their ability to distinguish between the bona fide and plagiarized responses.}
    \label{Proposedframework}
\end{figure}

The detection pipeline is summarized in Figure \ref{Proposedframework}. It consists of data segmentation, feature extraction, training classification models that use keystroke features, and testing those models to distinguish between bona fide and assisted submissions. We describe our assumptions and these components in the following sections. 

\subsubsection{Assumptions}
We define a plagiarism-detection system for identifying academic dishonesty under the following deployment assumptions:

\begin{itemize}
    \item Students complete assignments or examinations by typing their responses in a controlled environment that restricts advanced editing functionalities, including copy-and-paste, auto-completion, file uploads, or any other mechanisms that allow text input without generating observable keystrokes. Some widely deployed integrated development environments and classroom management platforms support such restrictions \cite{gdb}.
    
    \item The response platform can record keystroke data during the writing process, including key identifiers, \textit{KeyDown} and \textit{KeyUp} events, and timestamps. Students explicitly consent to this keystroke logging as part of their participation. While students with cheating intent may attempt to alter their typing behavior by slowing down or introducing artificial editing and revision patterns (as discussed in Section~\ref{sec:deception_threat_model}), we do not assume that students can spoof keystroke traces by directly injecting or manipulating the logging mechanism.
    
    \item In certain deployment scenarios, users complete a short reference typing task to capture \emph{fixed} keystroke patterns, which are subsequently used to evaluate user-specific detection models.
\end{itemize}

Under these assumptions, we consider plagiarism detection within controlled electronic assessment platforms that provide reliable keystroke logging.

\subsubsection{Keystroke modeling and classification}    
Over the past three decades, a wide range of techniques have been proposed for modeling and classifying keystroke dynamics \cite{TPAMIKeystroke1990, press1980authentication, shadman2023keystroke}. These approaches can be broadly grouped into statistical, machine learning (ML), and deep learning (DL) methods. Statistical approaches include Gaussian Mixture Models (GMM), Absolute Verifiers \cite{Gunetti2005KeystrokeAO}, Similarity Verifiers \cite{phoha2010methods}, and Instance-based Tail Area Density (ITAD) \cite{itad_ref0, itad_ref}, which operate directly on temporal features such as key hold times and inter-key intervals computed for individual keys and their combinations (e.g., pairs and triplets) \cite{wahab2023simple, shadman2023keystroke, kuruvilla2024spotting}. 

Machine learning classifiers, including Support Vector Machines (SVM), k-Nearest Neighbors (k-NN), Random Forests (RF), Naive Bayes, Gradient Boosting, Logistic Regression, and Multilayer Perceptrons (MLP), have also been widely used, typically trained on aggregated statistics derived from these temporal features \cite{TypingPhoneBTAS2016, shadman2023keystroke, teh2013survey}. More recently, deep learning methods such as Recurrent Neural Networks (RNN) and Convolutional Neural Networks (CNN), including architectures such as TypeNet, have been employed to automatically learn discriminative representations from keystroke data, resulting in improved accuracy and adaptability \cite{shadman2023keystroke, acien2021typenet, acien2022detection, keystrokesCNN, Sharma2023CNNKeystrokes}. 

In this work, we evaluate three representative classifiers: Light Gradient-Boosting Machine (LGBM), CatBoost, and TypeNet. The training procedures for these models, along with the rationale for their selection, are described below.

\subsubsection{ML-based Detectors Training}
Segmentation and feature extraction are used to transform raw keystroke streams into feature vectors for training ML-based detectors. Keystroke sequences are segmented using a sliding window of $1000$ keystrokes with an overlap of $300$ keystrokes. This windowing strategy captures local temporal patterns and provides robustness to noise and missing events, enabling deployment in online and real-time assessment settings.

Following prior work \cite{TypingPhoneBTAS2016}, we extract standard temporal features from each window, including dwell time (key hold time or unigrams) and flight time (key interval time or digraphs). In addition, we compute window-level features comprising the number of words, backspace count per sentence, average word length, nouns per sentence, verbs per sentence, modifiers per sentence, modals per sentence, named entities per sentence, lexical density, number of spelling mistakes, number of special characters per sentence, and keystrokes per burst. For keystroke timing features, we compute the mean values for keys and key pairs that occur more than once within a window.

Keystroke feature representations are inherently sparse, as not all keys or key pairs appear in every window. To address this, we retain only the $100$ most frequent unigram and digraph features, computed from the training data. The resulting feature vectors are used to train two gradient-boosting classifiers: Light Gradient Boosting Machine (LGBM) \cite{LGBM} and CatBoost \cite{CatBoost}.

\subsubsection{ML-based Detectors Training}
Segmentation and feature extraction are used to transform raw keystroke streams into feature vectors for training ML-based classifiers. Keystroke sequences are segmented using a sliding window of $1000$ keystrokes with an overlap of $300$ keystrokes. This windowing strategy captures local temporal patterns and provides robustness to noise and missing events, making it suitable for real-time detection in online examination settings.

Following prior studies such as \cite{TypingPhoneBTAS2016}, we extract standard temporal features from each window, including dwell time (key-hold time or unigrams) and flight time (key-interval time or digraphs). In addition, we compute window-level features comprising the number of words, backspace count per sentence, average word length, nouns per sentence, verbs per sentence, modifiers per sentence, modals per sentence, named entities per sentence, lexical density, number of spelling mistakes, number of special characters per sentence, and keystrokes per burst. For keystroke timing features, we compute the mean values for keys and key pairs that occur more than once within a window.

Keystroke data are inherently sparse, as not all keys or key pairs appear in every window. Accordingly, we retain only the $100$ most frequent unigram and digraph features, with frequencies computed from the training data. The resulting feature vectors are used to train two gradient-boosting classifiers: Light Gradient Boosting Machine (LGBM) \cite{LGBM} and CatBoost \cite{CatBoost}. These models are well-suited to the high-dimensional, heterogeneous feature space formed by sparse timing features and aggregated linguistic and editing markers, and can model nonlinear interactions relevant to distinguishing assisted from bona fide typing behavior.



\subsubsection{DL-based Detector Training}
We repurpose the TypeNet architecture \cite{acien2021typenet}, which is based on Long Short-Term Memory (LSTM) networks originally developed for keystroke-based user authentication. TypeNet models sequential keystroke data to capture temporal typing characteristics such as rhythm, timing, and key-hold duration. In its original formulation, TypeNet was trained using softmax, contrastive, and triplet loss functions, and required approximately $250$ keystrokes per user during training to achieve stable authentication performance \cite{acien2021typenet}.

We adapt TypeNet into a Siamese architecture to distinguish between bona fide and assisted typing behaviors, including transcription and paraphrasing. The modified architecture consists of: (1) two LSTM layers with $128$-dimensional outputs, each followed by batch normalization and \textit{tanh} activation; (2) a fully connected layer after each LSTM to project the representations into a shared $128$-dimensional embedding space; and (3) dropout layers after each LSTM with a dropout rate of $0.5$ and a recurrent dropout rate of $0.2$. We vary the sequence length \(M\) and batch size during training, exploring \(M \in [25, 500]\) and batch sizes from $32$ to $512$ (powers of two). Training is performed for up to $100$ epochs, with performance typically converging between $50$ and $100$ epochs. Optimization uses Adam with a learning rate in the range $[0.0001, 0.005]$, along with $L_2$ regularization.

Keystroke sequences are standardized to a fixed length \(M\) through padding or truncation, with \(M\) selected during hyperparameter tuning. Sequences with excessive Shift key usage (greater than $20$\%) or with effective lengths shorter than $50$\% of \(M\) are excluded. Each keystroke event is represented by three attributes: key action (KeyDown or KeyUp), key code, and timestamp. Timestamps are min--max normalized to the range $[0,1]$, keycodes are normalized by dividing by $255$, and key actions are binary encoded, with KeyDown mapped to $0$ and KeyUp to $1$. Pairs of sequences are formed from fixed and free typing conditions for Siamese training. Pairs from different typing conditions are labeled as $1$, while pairs from the same condition are labeled as $0$. A balanced pairing strategy is used to avoid class imbalance.

Unlike the original TypeNet, which uses Euclidean distance to compare embeddings, we employ cosine similarity within the loss formulation to measure embedding similarity. Cosine similarity is scale-invariant and bounded within $[-1,1]$, reducing sensitivity to variations in embedding magnitude caused by differences in typing speed, users, devices, or environments. In our preliminary experiments, cosine similarity resulted in more stable optimization behavior and consistently improved performance compared to Euclidean distance.

Training uses Binary Cross-Entropy loss computed over cosine similarity scores. The decision threshold for classifying a pair as bona fide or assisted is determined using the Equal Error Rate (EER) operating point on the ROC curve. This threshold is then used to compute Accuracy, $F_1$ score, False Acceptance Rate (FAR), and False Rejection Rate (FRR). Additional implementation details are available in our publicly released code.\footnote{https://github.com/ijcb-2024/keystroke-llm-plagiarism}

\subsection{Threat Modeling and Robustness Evaluation}
\subsubsection{Threat Model} \label{sec:deception_threat_model}
Building on the attack datasets described in Section~\ref{sec:attack_dataset}, we define a deception-based threat model to evaluate the robustness and generalization of both ML- and DL-based detectors. The threat model characterizes how a motivated adversary (e.g., a student) may attempt to evade a keystroke-dynamics-based plagiarism detection system. We specify the adversary’s \textit{goal}, \textit{knowledge}, and \textit{capabilities}.

\underline{Goal.}  
The adversary aims to submit a plagiarized or LLM-assisted response, denoted by $x_{2}$, together with a forged keystroke sequence $\hat{K}(x_{2})$ that is statistically consistent with bona fide human typing behavior. The attack is considered successful if the detector classifies the pair $(x_{2}, \hat{K}(x_{2}))$ as bona fide.

\underline{Knowledge.}  
We assume that the adversary has access to at least one authentic reference sample $(x_{1}, K(x_{1}))$, where $x_{1}$ is a text previously typed by the same user and $K(x_{1})$ denotes the corresponding keystroke sequence comprising key-down events, key-up events, and timestamps. This assumption reflects a realistic scenario in which students can access their own past typed content and associated typing behavior. The adversary does not have access to the internal architecture, parameters, or decision thresholds of the plagiarism detector.

\underline{Capabilities.}  
Given a reference sample $(x_{1}, K(x_{1}))$ and a new target text $x_{2}$, the adversary is capable of generating a forged keystroke sequence $\hat{K}(x_{2})$ by transferring timing characteristics from $K(x_{1})$ and aligning them with the characters of $x_{2}$. Formally, this process is represented as
\[
\hat{K}(x_{2}) = \mathcal{G}(x_{1}, K(x_{1}), x_{2}),
\]
where $\mathcal{G}$ denotes a keystroke generator. Prior work on keystroke synthesis \cite{GONZALEZ2023100454} produces realistic timing distributions but does not explicitly enforce alignment between generated timings and coherent textual content. To address this limitation, we design a generator that reuses unigram and digraph timing statistics from $K(x_{1})$ to map onto the characters of $x_{2}$, with missing values interpolated using smoothed random perturbations calibrated to the empirical timing distribution. This generator is used to construct the attack datasets described in Section~\ref{sec:attack_dataset}.

\underline{Outcome.}  
The detector receives the input pair $(x_{2}, \hat{K}(x_{2}))$. If this pair is classified as bona fide, the deception attempt succeeds. This threat model enables systematic evaluation of detector robustness in realistic scenarios where keystroke dynamics are synthetically forged while the submitted text remains semantically valid.

\subsubsection{Vulnerability Analysis}
Let $(x_{1}, K(x_{1}))$ denote a student’s authentic text and its associated keystroke dynamics, and let $x_{2}$ denote a plagiarized or LLM-assisted text. Using the deception generator $\mathcal{G}$ defined in Section~\ref{sec:attack_dataset}, we synthesize forged keystroke sequences
\[
\hat{K}(x_{2}) = \mathcal{G}(x_{1}, K(x_{1}), x_{2}),
\]
which simulate deception attempts under the defined threat model.

Each detector $f_{\theta}$ is evaluated on input pairs $(x_{2}, \hat{K}(x_{2}))$ to assess its vulnerability:
\[
f_{\theta}(x_{2}, \hat{K}(x_{2})) \rightarrow y',
\]
where $y' = 0$ indicates that a plagiarized instance is misclassified as bona fide. Under our label convention (see \ref{evaluation_metrics}), this corresponds to a false rejection of plagiarism (i.e., a false negative), and is quantified using the False Reject Rate (FRR).

\subsubsection{Countermeasure}
Experiments conducted under the proposed deception-based threat model show substantial performance degradation in ML-based detectors, indicating their vulnerability to forged keystroke dynamics (Table~\ref{tab:attack}). To mitigate this vulnerability, we employ an adversarial training strategy that explicitly incorporates forged keystroke sequences into the training process.

The adversarial training procedure augments the original training data with synthetically generated deception samples. For each bona fide training instance $(x_i, K(x_i))$, we generate a corresponding forged sample
\[
(x_i', \hat{K}(x_i')) = (x_i', \mathcal{G}(x_i, K(x_i), x_i'))
\]
using the keystroke generator $\mathcal{G}$ defined in the threat model. The resulting adversarially augmented dataset is given by
\[
D_{\mathrm{adv}} = D_{\mathrm{train}} \cup \{(x_i', \hat{K}(x_i'), y_i' = 0)\}.
\]

Training on $D_{\mathrm{adv}}$ enables the detector to learn decision boundaries that better separate genuine and forged keystroke dynamics, thereby improving robustness to deception-based attacks (Figure~\ref{fig:decision_boundary}).

As with the threat model and vulnerability analysis, the effectiveness of this countermeasure is bounded by the stated assumptions about the adversary's knowledge and capabilities. More adaptive or learned generators, such as those designed to mimic user-specific latency profiles conditioned on text, may pose stronger threats and are left for future investigation.

\subsection{Performance Evaluation}
\subsubsection{Evaluation Metrics}
\label{evaluation_metrics}
We report the False Acceptance Rate (FAR), False Rejection Rate (FRR), and $F_1$ score for all plagiarism detectors. In our formulation, \underline{plagiarism is treated as the positive class}, while bona fide writing is treated as the negative class \cite{Canyakan_2025, ai_academic_integrity}. Under this convention, FAR corresponds to the proportion of plagiarized responses that are incorrectly classified as bona fide (missed plagiarism), whereas FRR corresponds to the proportion of bona fide responses that are incorrectly classified as plagiarized (false accusations).

Reporting both FAR and FRR provides a comprehensive view of detector behavior by capturing errors associated with both missed plagiarism and wrongful accusations, which is particularly important in high-stakes educational assessment settings.

The $F_1$ score is defined as the harmonic mean of precision and recall, where precision measures the proportion of correctly identified plagiarized responses among all responses flagged as plagiarized, and recall measures the proportion of actual plagiarized responses that are correctly detected. The $F_1$ score provides a balanced summary of detection performance by jointly accounting for both types of errors.

\subsubsection{Evaluation Under Invariants}
Evaluating the proposed detectors under practical invariants is necessary because typing behavior can vary across users, keyboards, and writing contexts, among other factors. Exhaustively testing all possible variations is impractical due to data availability constraints. Based on the four datasets used in this study, we evaluate detector performance under user-, keyboard-, context-, and dataset-specific conditions, as well as their corresponding agnostic settings. The construction of these evaluation scenarios is described below.

\textit{User-specific} evaluation trains and tests a separate model for each user using an $80$--$20$ split of that user’s data, with no overlap between training and testing sequences. In contrast, \textit{user-agnostic} evaluation assesses generalization across users by splitting the full user population into disjoint training and test sets, each comprising $80\%$ and $20\%$ of users, respectively.

In the \textit{keyboard-specific} setting, models are trained and tested on data collected using the same keyboard type, again using an $80$--$20$ split with distinct sequences. The \textit{keyboard-agnostic} setting trains models on data from three keyboard types and evaluates them on a fourth, unseen keyboard. This evaluation uses the Buffalo dataset, which explicitly annotates keyboard types and distinguishes between free and fixed typing sequences. For reference, the Lenovo, HP, Microsoft, and Apple Bluetooth keyboards are denoted $K_{0}$, $K_{1}$, $K_{2}$, and $K_{3}$, respectively.

\textit{Context-specific} evaluation trains and tests models on sequences drawn from the same writing topic, ensuring topic homogeneity within each split. An $80$--$20$ split is applied within each context, with no sequence overlap. Conversely, \textit{context-agnostic} evaluation trains models on data from two contexts and tests them on a third, previously unseen context, thereby assessing robustness to topical variation. This evaluation is performed using the SBU dataset, which provides three distinct contexts: Gay Marriage (GM), Gun Control (GC), and Restaurant Feedback (RF).

Finally, \textit{dataset-specific} evaluation trains and tests models on the same dataset using non-overlapping splits, thereby capturing performance under consistent data-collection conditions. In contrast, \textit{dataset-agnostic} evaluation trains models on combinations of datasets and tests them on different datasets or dataset mixtures, assessing cross-dataset generalization. Together, these evaluation settings provide a comprehensive view of detector robustness across different users, devices, contexts, and data-collection environments.

\subsubsection{Benchmarking DetectGPT}
We benchmark DetectGPT \cite{mitchell2023detectgpt}, a text-intrinsic plagiarism detector, on the datasets used in this study. The underlying \textit{EleutherAI/gpt-neo-2.7B} model has a maximum context length of 2048 tokens, which makes response-level evaluation infeasible for datasets in which a substantial portion of student responses exceed this limit.

To ensure compatibility with this constraint, we apply DetectGPT at two granularities: \textit{window-level} and \textit{sentence-level}. For window-level evaluation, responses are segmented into sliding windows of $1000$ tokens with a $300$-token overlap, each of which falls within the model’s context budget. Each window is independently scored by DetectGPT, and predictions are aggregated across windows to obtain a response-level decision. For sentence-level evaluation, individual sentences are treated as independent units of analysis, enabling assessment of DetectGPT’s performance on shorter text spans.

This evaluation strategy allows DetectGPT to be applied consistently across all datasets without exceeding its context-length limitations, while providing complementary insights into its behavior at different levels of textual granularity.

\subsubsection{Benchmarking LLaMA-3.3-70B-Instruct}
To evaluate the ability of pretrained large language models (LLMs) to distinguish between human-authored and AI-assisted text, we benchmark LLaMA-3.3-70B-Instruct across two learning paradigms: \textit{zero-shot} and \textit{few-shot}. Experiments are conducted across two input configurations—\textit{response-level} and \textit{window-level}—and two input formats—\textit{raw} and \textit{preprocessed} text.

In the \textit{zero-shot} setting, the model relies solely on its pretrained knowledge and task-specific prompts to classify text as \textit{Original/Bona fide} or \textit{Copied/Assisted}. In the \textit{few-shot} setting, the model is provided with six labeled examples covering transcription and paraphrasing scenarios, supplying task-specific context to guide classification.

In the \textit{response-level} configuration, the full text response is provided as input, enabling the model to leverage global linguistic and structural cues. Prompts in this setting operate on complete responses (e.g., \textit{“Classify the following response as Original/Bona fide or Copied/Assisted (transcribed or paraphrased)”}). In contrast, the \textit{window-level} configuration segments responses into fixed-length text fragments (approximately 300 characters), allowing localized analysis and simulating partial-observation scenarios. Prompts are correspondingly adapted for segment-level classification.

To assess the impact of input preparation, we evaluate two text formats. The \textit{raw} format preserves the original student text, including typographical errors, formatting inconsistencies, and other artifacts of human typing. The \textit{preprocessed} format standardizes text by normalizing capitalization, correcting spelling and punctuation, removing extraneous whitespace, and enforcing consistent formatting. This comparison isolates the effect of surface-level noise on detection performance.

Prompts are designed to enforce binary classification without explanatory output. For example, in the zero-shot setting, the model is instructed to output only \textit{Original/Bona fide} or \textit{Copied/Assisted}, with explicit definitions of both categories.

Preliminary experiments were conducted with smaller LLaMA variants (LLaMA-3.1-8B-Instruct and LLaMA-2-13B) and ChatGPT-3.5. These models did not achieve performance beyond chance on this task. Although ChatGPT-4.0 showed stronger performance, it was excluded due to prohibitive API costs. LLaMA-3.3-70B-Instruct was therefore selected as a practical balance between performance and cost. Detailed results across learning paradigms, input configurations, and text formats are reported in Section~\ref{LLaMAResults}.

\subsubsection{Benchmarking Human Evaluators}
In addition to benchmarking automated text-intrinsic detectors, we evaluate human performance in distinguishing between bona fide and AI-assisted plagiarism. This study used a web-based survey comprising four evaluation scenarios: response-level transcription, response-level paraphrasing, window-level transcription, and window-level paraphrasing.

In each survey, $10$ text samples were randomly selected from the corresponding dataset and presented to participants for classification as either \textit{human-generated} or \textit{AI-assisted} text. Randomization ensured that participants received distinct samples and minimized selection bias. Response-level surveys presented complete text responses, allowing participants to assess global linguistic and stylistic characteristics, whereas window-level surveys presented fixed-length text segments of approximately $300$ characters, emphasizing localized cues.

A total of $44$ human evaluators were recruited from university campuses, including graduate students, researchers, and faculty members. Participants reported their prior experience with large language models and generative AI tools. Based on these responses, evaluators were categorized into two groups: \textit{experts} ($14$ participants) and \textit{non-experts} ($30$ participants). This stratification enables analysis of the effect of domain familiarity on human detection performance.

Each participant completed all four surveys, resulting in $40$ classification decisions per participant ($10$ per scenario). Aggregate results and comparative analysis are presented in Section~\ref{HumanPerfEval}.

\section{Results and Discussion}
\label{sec:results_discussions}
This section reports the performance of the proposed keystroke-based detectors, the pretrained LLM LLaMA-3.3-70B-Instruct, human evaluators, and the models' robustness to deception-based attacks. 

Among ML-based detectors, LightGBM and CatBoost achieved the highest $F_1$ scores in structured evaluation settings, with performance ranging from $97.2\%$ to $99.1\%$ in keyboard-specific scenarios and from $97.5\%$ to $98.2\%$ in user-specific scenarios. However, their performance degraded substantially on paraphrased data, with $F_1$ scores dropping to $75.7\%$--$76.4\%$. In contrast, the DL-based TypeNet model demonstrated stronger robustness to paraphrasing, achieving an $F_1$ score of $86.9\%$ in paraphrased scenarios.

Pretrained LLMs exhibited highly variable performance across tasks, learning paradigms, and input preparation strategies. For transcription detection, input preprocessing substantially improved performance, with $F_1$ scores increasing from $5.3\%$--$76.1\%$ under zero-shot learning and from $48.4\%$--$83.1\%$ under few-shot learning. Performance on paraphrasing tasks was less consistent, with $F_1$ scores ranging from $7.5\%$--$45.0\%$ in zero-shot settings and $21.8\%$--$72.5\%$ in few-shot settings.

Human evaluators exhibited considerable variability in performance. Overall, $F_1$ scores ranged from approximately $50\%$, corresponding to chance-level accuracy, to above $70\%$ for expert participants. This variability highlights the inherent difficulty of reliably identifying AI-assisted text, particularly in paraphrasing scenarios.

It is important to note that the Buffalo and SBU datasets differ fundamentally from the IIITD-BU-Transcribed and IIITD-BU-Paraphrased datasets. The Buffalo and SBU datasets consist of structured typing tasks without LLM-assisted transcription or paraphrasing, whereas the IIITD-BU datasets are explicitly designed to capture LLM-assisted plagiarism. The strong performance of ML-based detectors on the Buffalo and SBU datasets reflects their effectiveness in structured environments, while TypeNet's improved performance on IIITD-BU-Paraphrased highlights its ability to handle more complex, linguistically nuanced plagiarism scenarios. The following subsections provide a detailed analysis of these results.

\begin{table*}[htp]
\centering
\caption{Performance comparison of ML-based detectors (LGBM, CatBoost) and TypeNet across user-, keyboard-, context-, and dataset-specific scenarios using $F_1$ score. ML models achieve higher $F_1$ in user-specific (97.5--98.2) and keyboard-specific (97.2--99.1) settings, while TypeNet performs better in context-specific (84.5--85.7) and paraphrased dataset evaluations (86.9 on IIITD-BU-Paraphrased).}

\label{tab:scenario-specific-performances}
\begin{tabular}{|cc|ccccccccc|}
\hline
\multicolumn{2}{|c|}{\textbf{Dataset and splits}}                 & \multicolumn{3}{c|}{\textbf{LGBM}}                                                                       & \multicolumn{3}{c|}{\textbf{CatBoost}}                                                                   & \multicolumn{3}{c|}{\textbf{TypeNet}}                                               \\ \hline
\multicolumn{1}{|c|}{\textbf{Train}}      & \textbf{Test}         & \multicolumn{1}{c|}{\textbf{F1}} & \multicolumn{1}{c|}{\textbf{FAR}} & \multicolumn{1}{c|}{\textbf{FRR}} & \multicolumn{1}{c|}{\textbf{F1}} & \multicolumn{1}{c|}{\textbf{FAR}} & \multicolumn{1}{c|}{\textbf{FRR}} & \multicolumn{1}{c|}{\textbf{F1}} & \multicolumn{1}{c|}{\textbf{FAR}} & \textbf{FRR} \\ \hline
\multicolumn{1}{|l|}{}                    & \multicolumn{1}{l|}{} & \multicolumn{9}{c|}{\textbf{User-specific (all four datasets)}}                                                                                                                                                                                                                                                               \\ \hline
\multicolumn{2}{|c|}{\textbf{Combined (80-20)}}                   & \multicolumn{1}{c|}{97.5}       & \multicolumn{1}{c|}{4.9}         & \multicolumn{1}{c|}{0}            & \multicolumn{1}{c|}{98.2}       & \multicolumn{1}{c|}{2.53}         & \multicolumn{1}{c|}{0.98}         & \multicolumn{1}{c|}{77.4}       & \multicolumn{1}{c|}{31.7}        & 23.1        \\ \hline
\multicolumn{1}{|l|}{}                    & \multicolumn{1}{l|}{} & \multicolumn{9}{c|}{\textbf{Keyboard-specific (Buffalo dataset only)}}                                                                                                                                                                                                                                                           \\ \hline
\multicolumn{1}{|c|}{K0}                  & K0                    & \multicolumn{1}{c|}{99.1}        & \multicolumn{1}{c|}{3.6}         & \multicolumn{1}{c|}{0.4}         & \multicolumn{1}{c|}{98.8}        & \multicolumn{1}{c|}{4.1}         & \multicolumn{1}{c|}{0.7}         & \multicolumn{1}{c|}{83.5}       & \multicolumn{1}{c|}{25.4}        & 19.8        \\ \hline
\multicolumn{1}{|c|}{K1}                  & K1                    & \multicolumn{1}{c|}{98}          & \multicolumn{1}{c|}{7.1}         & \multicolumn{1}{c|}{1.1}          & \multicolumn{1}{c|}{98.4}       & \multicolumn{1}{c|}{4.4}         & \multicolumn{1}{c|}{1.5}         & \multicolumn{1}{c|}{73.3}       & \multicolumn{1}{c|}{32.6}         & 30.16        \\ \hline
\multicolumn{1}{|c|}{K2}                  & K2                    & \multicolumn{1}{c|}{97.2}       & \multicolumn{1}{c|}{6.3}         & \multicolumn{1}{c|}{2.8}         & \multicolumn{1}{c|}{98.1}       & \multicolumn{1}{c|}{6.3}         & \multicolumn{1}{c|}{1.2}         & \multicolumn{1}{c|}{76.6}       & \multicolumn{1}{c|}{32.1}        & 25.1        \\ \hline
\multicolumn{1}{|c|}{K3}                  & K3                    & \multicolumn{1}{c|}{98.5}       & \multicolumn{1}{c|}{3.5}         & \multicolumn{1}{c|}{1.5}         & \multicolumn{1}{c|}{98.0}       & \multicolumn{1}{c|}{5.2}         & \multicolumn{1}{c|}{1.9}         & \multicolumn{1}{c|}{78.9}       & \multicolumn{1}{c|}{29.1}        & 24.2       \\ \hline
\multicolumn{1}{|l|}{}                    & \multicolumn{1}{l|}{} & \multicolumn{9}{c|}{\textbf{Context-specific (SBU dataset only)}}                                                                                                                                                                                                                                                            \\ \hline
\multicolumn{1}{|c|}{GM}                  & GM                    & \multicolumn{1}{c|}{80.2}       & \multicolumn{1}{c|}{22.7}        & \multicolumn{1}{c|}{16.5}        & \multicolumn{1}{c|}{81.6}       & \multicolumn{1}{c|}{20.2}        & \multicolumn{1}{c|}{15.8}        & \multicolumn{1}{c|}{85.7}       & \multicolumn{1}{c|}{21.5}        & 17.7        \\ \hline
\multicolumn{1}{|c|}{GC}                  & GC                    & \multicolumn{1}{c|}{80.7}       & \multicolumn{1}{c|}{22.1}        & \multicolumn{1}{c|}{16.0}        & \multicolumn{1}{c|}{80.1}       & \multicolumn{1}{c|}{23.6}        & \multicolumn{1}{c|}{16.0}        & \multicolumn{1}{c|}{85.2}       & \multicolumn{1}{c|}{22.9}        & 21.9        \\ \hline
\multicolumn{1}{|c|}{RF}                  & RF                    & \multicolumn{1}{c|}{78.9}       & \multicolumn{1}{c|}{23.1}        & \multicolumn{1}{c|}{18.5}        & \multicolumn{1}{c|}{79.8}       & \multicolumn{1}{c|}{24.0}        & \multicolumn{1}{c|}{16.3}         & \multicolumn{1}{c|}{84.5}       & \multicolumn{1}{c|}{30.5}        & 18.4        \\ \hline
\multicolumn{1}{|l|}{}                    & \multicolumn{1}{l|}{} & \multicolumn{9}{c|}{\textbf{Dataset-specific}}                                                                                                                                                                                                                                                            \\ \hline
\multicolumn{1}{|c|}{SBU}                 & SBU                   & \multicolumn{1}{c|}{86.1}        & \multicolumn{1}{c|}{9.6}         & \multicolumn{1}{c|}{13.2}        & \multicolumn{1}{c|}{84.8}       & \multicolumn{1}{c|}{11}           & \multicolumn{1}{c|}{13.9}        & \multicolumn{1}{c|}{86.1}        & \multicolumn{1}{c|}{13.2}        & 9.6         \\ \hline
\multicolumn{1}{|c|}{Buffalo}             & Buffalo               & \multicolumn{1}{c|}{98.9}       & \multicolumn{1}{c|}{1.1}         & \multicolumn{1}{c|}{2.8}         & \multicolumn{1}{c|}{98.7}       & \multicolumn{1}{c|}{4.1}         & \multicolumn{1}{c|}{0.9}         & \multicolumn{1}{c|}{71.9}       & \multicolumn{1}{c|}{31.8}        & 33.6        \\ \hline
\multicolumn{1}{|c|}{(IIITD-BU-Transcribed)}         & (IIITD-BU-Transcribed)           & \multicolumn{1}{c|}{90.9}       & \multicolumn{1}{c|}{4.0}         & \multicolumn{1}{c|}{12.2}         & \multicolumn{1}{c|}{90.3}       & \multicolumn{1}{c|}{3.6}         & \multicolumn{1}{c|}{13.8}        & \multicolumn{1}{c|}{82.2}       & \multicolumn{1}{c|}{22.6}         & 28.2         \\ \hline
\multicolumn{1}{|c|}{(IIITD-BU-Paraphrased)}         & (IIITD-BU-Paraphrased)           & \multicolumn{1}{c|}{76.4}       & \multicolumn{1}{c|}{15.9}        & \multicolumn{1}{c|}{26.3}         & \multicolumn{1}{c|}{75.7}       & \multicolumn{1}{c|}{16.3}        & \multicolumn{1}{c|}{27.1}        & \multicolumn{1}{c|}{86.9}       & \multicolumn{1}{c|}{17.5}        & 17.4        \\ \hline
\multicolumn{1}{|c|}{(IIITD-BU-Combined)} & (IIITD-BU-Combined)   & \multicolumn{1}{c|}{83.8}       & \multicolumn{1}{c|}{24.8}        & \multicolumn{1}{c|}{13.5}         & \multicolumn{1}{c|}{82.7}        & \multicolumn{1}{c|}{25.9}        & \multicolumn{1}{c|}{14.9}        & \multicolumn{1}{c|}{75.1}        & \multicolumn{1}{c|}{28.3}        & 29.2        \\ \hline
\end{tabular}
\end{table*}

\begin{table*}[htp]
\centering
\caption{Performance of ML-based detectors (LGBM, CatBoost) and TypeNet under user-, keyboard-, context-, and dataset-agnostic scenarios using $F_1$ score. ML models show strong generalization in user-agnostic (97.0--97.2) and keyboard-agnostic (98.0--98.5) settings, while TypeNet attains lower but improved performance (68.5--78.8). In context- and dataset-agnostic evaluations, ML models remain more stable, whereas TypeNet exhibits lower and more variable performance.}
\label{tab:scenario-agnostic-performances}
\begin{tabular}{|cc|ccccccccc|}
\hline
\multicolumn{2}{|c|}{\textbf{Dataset and splits}}                             & \multicolumn{3}{c|}{\textbf{LGBM}}                                                                       & \multicolumn{3}{c|}{\textbf{CatBoost}}                                                                   & \multicolumn{3}{c|}{\textbf{TypeNet}}                                               \\ \hline
\multicolumn{1}{|c|}{\textbf{Train}}             & \textbf{Test}              & \multicolumn{1}{c|}{\textbf{F1}} & \multicolumn{1}{c|}{\textbf{FAR}} & \multicolumn{1}{c|}{\textbf{FRR}} & \multicolumn{1}{c|}{\textbf{F1}} & \multicolumn{1}{c|}{\textbf{FAR}} & \multicolumn{1}{c|}{\textbf{FRR}} & \multicolumn{1}{c|}{\textbf{F1}} & \multicolumn{1}{c|}{\textbf{FAR}} & \textbf{FRR} \\ \hline
\multicolumn{1}{|l|}{}                           & \multicolumn{1}{l|}{}      & \multicolumn{9}{c|}{\textbf{User-agnostic (all four datasets)}}                                                                                                                                                                                                                                                               \\ \hline
\multicolumn{2}{|c|}{\textbf{Combined (80-20)}}                               & \multicolumn{1}{c|}{97.0}       & \multicolumn{1}{c|}{3.4}          & \multicolumn{1}{c|}{2.3}          & \multicolumn{1}{c|}{97.2}       & \multicolumn{1}{c|}{3.4}         & \multicolumn{1}{c|}{1.9}         & \multicolumn{1}{c|}{68.5}        & \multicolumn{1}{c|}{40.8}        & 43.81        \\ \hline
\multicolumn{1}{|l|}{}                           & \multicolumn{1}{l|}{}      & \multicolumn{9}{c|}{\textbf{Keyboard-agnostic (Buffalo dataset only)}}                                                                                                                                                                                                                                                           \\ \hline
\multicolumn{1}{|c|}{\textbf{K0,1,2}}            & \textbf{K3}                & \multicolumn{1}{c|}{98.0}       & \multicolumn{1}{c|}{6.6}         & \multicolumn{1}{c|}{1.2}         & \multicolumn{1}{c|}{98.0}       & \multicolumn{1}{c|}{6.9}         & \multicolumn{1}{c|}{1.2}         & \multicolumn{1}{c|}{75.9}       & \multicolumn{1}{c|}{28.0}        & 29.0        \\ \hline
\multicolumn{1}{|c|}{\textbf{K0,1,3}}            & \textbf{K2}                & \multicolumn{1}{c|}{98.5}       & \multicolumn{1}{c|}{2.4}         & \multicolumn{1}{c|}{1.9}         & \multicolumn{1}{c|}{98.3}       & \multicolumn{1}{c|}{3.2}         & \multicolumn{1}{c|}{1.9}         & \multicolumn{1}{c|}{78.1}       & \multicolumn{1}{c|}{27.5}         & 28.4        \\ \hline
\multicolumn{1}{|c|}{\textbf{K0,2,3}}            & \textbf{K1}                & \multicolumn{1}{c|}{98.3}       & \multicolumn{1}{c|}{3.1}         & \multicolumn{1}{c|}{2.1}         & \multicolumn{1}{c|}{98.3}       & \multicolumn{1}{c|}{3.4}         & \multicolumn{1}{c|}{1.9}          & \multicolumn{1}{c|}{78.8}       & \multicolumn{1}{c|}{24.0}        & 28.4        \\ \hline
\multicolumn{1}{|c|}{\textbf{K1,2,3}}            & \textbf{K0}                & \multicolumn{1}{c|}{98.1}       & \multicolumn{1}{c|}{6.3}          & \multicolumn{1}{c|}{1.3}          & \multicolumn{1}{c|}{98.2}       & \multicolumn{1}{c|}{6.2}         & \multicolumn{1}{c|}{1.0}         & \multicolumn{1}{c|}{76.9}       & \multicolumn{1}{c|}{29.2}        & 25.0        \\ \hline
\multicolumn{1}{|l|}{}                           & \multicolumn{1}{l|}{}      & \multicolumn{9}{c|}{\textbf{Context-agnostic (SBU dataset only)}}                                                                                                                                                                                                                                                            \\ \hline
\multicolumn{1}{|c|}{\textbf{(GM,RF)}}           & \textbf{GC}                & \multicolumn{1}{c|}{81.1}       & \multicolumn{1}{c|}{22.0}        & \multicolumn{1}{c|}{15.4}        & \multicolumn{1}{c|}{81.1}       & \multicolumn{1}{c|}{21.9}        & \multicolumn{1}{c|}{15.5}        & \multicolumn{1}{c|}{67.0}       & \multicolumn{1}{c|}{35.6}        & 24.9        \\ \hline
\multicolumn{1}{|c|}{\textbf{(GC,RF)}}           & \textbf{GM}                & \multicolumn{1}{c|}{82.4}        & \multicolumn{1}{c|}{22.2}        & \multicolumn{1}{c|}{13.1}        & \multicolumn{1}{c|}{82.3}       & \multicolumn{1}{c|}{22.6}        & \multicolumn{1}{c|}{12.9}        & \multicolumn{1}{c|}{76.3}        & \multicolumn{1}{c|}{33.1}         & 22.3        \\ \hline
\multicolumn{1}{|c|}{\textbf{(GM,GC)}}           & \textbf{RF}                & \multicolumn{1}{c|}{80.2}       & \multicolumn{1}{c|}{27.3}        & \multicolumn{1}{c|}{13.1}        & \multicolumn{1}{c|}{80.7}       & \multicolumn{1}{c|}{26.0}        & \multicolumn{1}{c|}{13.2}        & \multicolumn{1}{c|}{70.2}       & \multicolumn{1}{c|}{39.7}        & 34.7        \\ \hline
\multicolumn{1}{|l|}{}                           & \multicolumn{1}{l|}{}      & \multicolumn{9}{c|}{\textbf{Dataset-agnostic}}                                                                                                                                                                                                                                                            \\ \hline
\multicolumn{1}{|c|}{\textbf{SBU, (IIITD-BU-Comb.)}}     & \textbf{Buffalo}           & \multicolumn{1}{c|}{75.7}       & \multicolumn{1}{c|}{12.4}        & \multicolumn{1}{c|}{35.5}        & \multicolumn{1}{c|}{69.9}       & \multicolumn{1}{c|}{24.5}         & \multicolumn{1}{c|}{18.7}         & \multicolumn{1}{c|}{68.7}       & \multicolumn{1}{c|}{42.1}        & 29.2        \\ \hline
\multicolumn{1}{|c|}{\textbf{SBU, Buffalo}}      & \textbf{(IIITD-BU-Comb.)}          & \multicolumn{1}{c|}{72.0}        & \multicolumn{1}{c|}{94.9}        & \multicolumn{1}{c|}{1.0}        & \multicolumn{1}{c|}{71.6}       & \multicolumn{1}{c|}{88.8}         & \multicolumn{1}{c|}{4.4}         & \multicolumn{1}{c|}{71.1}       & \multicolumn{1}{c|}{40.9}        & 27.2        \\ \hline
\multicolumn{1}{|c|}{\textbf{(IIITD-BU-Comb.), Buffalo}} & \textbf{SBU}               & \multicolumn{1}{c|}{84.1}       & \multicolumn{1}{c|}{4.2}          & \multicolumn{1}{c|}{22.3}         & \multicolumn{1}{c|}{87.8}       & \multicolumn{1}{c|}{2.7}         & \multicolumn{1}{c|}{20.2}         & \multicolumn{1}{c|}{65.9}        & \multicolumn{1}{c|}{42.1}        & 40.0        \\ \hline
\multicolumn{1}{|c|}{\textbf{SBU}}               & \textbf{(IIITD-BU-Comb.), Buffalo} & \multicolumn{1}{c|}{82.5}        & \multicolumn{1}{c|}{26}         & \multicolumn{1}{c|}{1.3}           & \multicolumn{1}{c|}{82.5}       & \multicolumn{1}{c|}{25.3}         & \multicolumn{1}{c|}{2.1}         & \multicolumn{1}{c|}{54.2}       & \multicolumn{1}{c|}{44.6}        & 38.3        \\ \hline
\multicolumn{1}{|c|}{\textbf{(IIITD-BU-Combined)}}          & \textbf{SBU, Buffalo}      & \multicolumn{1}{c|}{65.0}       & \multicolumn{1}{c|}{59.1}        & \multicolumn{1}{c|}{19.7}        & \multicolumn{1}{c|}{62.7}        & \multicolumn{1}{c|}{75.8}         & \multicolumn{1}{c|}{15.1}        & \multicolumn{1}{c|}{66.3}        & \multicolumn{1}{c|}{42.9}        & 33.6        \\ \hline
\multicolumn{1}{|c|}{\textbf{Buffalo}}           & \textbf{SBU, (IIITD-BU-Comb.)}     & \multicolumn{1}{c|}{73.0}       & \multicolumn{1}{c|}{55.9}        & \multicolumn{1}{c|}{18.0}         & \multicolumn{1}{c|}{64.8}       & \multicolumn{1}{c|}{52.0}         & \multicolumn{1}{c|}{18.4}         & \multicolumn{1}{c|}{62.9}       & \multicolumn{1}{c|}{43.4}        & 33.79        \\ \hline
\end{tabular}
\end{table*}

\subsection{Performance of the Proposed Models}
\subsubsection{Condition-Specific Performance}
We evaluate the performance of the ML-based detectors (LGBM and CatBoost) and the DL-based TypeNet across user-specific, keyboard-specific, context-specific, and dataset-specific scenarios. The results reveal clear condition-dependent strengths and limitations for each model.

\textit{User-specific scenarios.}
In user-specific evaluations, LGBM and CatBoost achieve consistently high performance, with $F_1$ scores ranging from $97.5\%$ to $98.2\%$, accompanied by minimal FAR and near-zero FRR. This indicates strong discrimination capability when training and testing data are drawn from the same user. In contrast, TypeNet performs substantially worse, achieving a maximum $F_1$ score of $77.4\%$ with FAR and FRR of $31.7\%$ and $23.1\%$, respectively. These results indicate that TypeNet is ineffective in highly personalized settings with limited per-user training data. Notably, our methodology does not rely on explicit user-specific calibration; rather, we evaluate both user-specific and user-agnostic settings and observe comparable performance across them.

\textit{Keyboard-specific scenarios.}
In keyboard-specific evaluations using the Buffalo dataset, ML-based detectors again demonstrate strong performance, achieving $F_1$ scores between $97.2\%$ and $99.1\%$ across different keyboard types. Both models maintain low error rates, with FAR values below $7.1\%$, and CatBoost exhibits slightly improved robustness to FAR. TypeNet shows moderate improvement over user-specific scenarios, with $F_1$ scores ranging from $73.3\%$ to $83.5\%$, but continues to underperform ML-based detectors. This performance gap suggests that TypeNet is less effective at capturing fine-grained keyboard-dependent variations.

\textit{Context-specific scenarios.}
In context-specific evaluations, TypeNet outperforms the ML-based detectors, achieving $F_1$ scores between $84.5\%$ and $85.7\%$, compared to $78.9\%$ to $81.6\%$ for LGBM and CatBoost. These results indicate that TypeNet benefits from training and testing under consistent topical contexts, where its sequential modeling better captures context-dependent typing behavior. While ML-based detectors remain competitive, their reliance on aggregated feature representations appears less effective in capturing contextual variability.

\textit{Dataset-specific scenarios.}
Dataset-specific evaluations show divergent behavior across datasets. ML-based detectors perform best on the Buffalo dataset, achieving a maximum $F_1$ score of $98.9\%$ with minimal FAR and FRR, reflecting their effectiveness in structured typing environments. In contrast, TypeNet achieves superior performance on the IIITD-BU-Paraphrased dataset, with an $F_1$ score of $86.9\%$, outperforming LGBM and CatBoost, which achieve $F_1$ scores of $75.7\%$ to $76.4\%$. This result highlights TypeNet’s relative advantage in handling paraphrased and linguistically complex inputs.

Overall, the results demonstrate that ML-based detectors excel in structured conditions such as user- and keyboard-specific scenarios, consistently achieving high $F_1$ scores with low error rates. Conversely, TypeNet shows greater robustness in context-rich, paraphrasing-heavy scenarios, where sequential modeling of keystroke dynamics provides an advantage. These findings indicate that no single model dominates across all conditions, underscoring the importance of selecting detection approaches based on the target deployment scenario. These results suggest that hybrid approaches combining feature-based ML models with sequence-based DL models could provide improved robustness across diverse deployment scenarios. We leave the exploration of such hybrid designs for future work.

\begin{table*}[htp]
\caption{Performance of LLaMA-3.3-70B-Instruct on the proposed datasets for transcription and paraphrasing tasks under zero-shot and few-shot learning paradigms, measured using $F_1$ score. For transcription tasks, input preprocessing improves performance substantially: response-level zero-shot $F_1$ increases from $5.3\%$ to $76.1\%$, and response-level few-shot $F_1$ increases from $48.4\%$ to $83.1\%$. For window-level paraphrasing, preprocessing increases zero-shot $F_1$ from $7.5\%$ to $45.0\%$ and few-shot $F_1$ from $21.8\%$ to $72.5\%$. In contrast, for response-level paraphrasing under zero-shot learning, preprocessing results in a decrease in $F_1$ score from $56.3\%$ to $41.1\%$.}

\label{tab:LLaMA-Instruct-Perf}
\centering
\begin{tabular}{|cc|rrrrrr|rrrrrr|}
\hline
\multicolumn{2}{|c|}{\multirow{3}{*}{\textbf{Dataset}}}                                                          & \multicolumn{6}{c|}{\textbf{Unprocessed (IIITD-BU-Combined)}}                                                                                                                                                                           & \multicolumn{6}{c|}{\textbf{Processed (IIITD-BU-Combined)}}                                                                                                                                                                             \\ \cline{3-14} 
\multicolumn{2}{|c|}{}                                                                                           & \multicolumn{3}{c|}{\textbf{Zero-shot learning}}                                                         & \multicolumn{3}{c|}{\textbf{Few-shot learning}}                                                          & \multicolumn{3}{c|}{\textbf{Zero-shot learning}}                                                         & \multicolumn{3}{c|}{\textbf{Few-shot learning}}                                                          \\ \cline{3-14} 
\multicolumn{2}{|c|}{}                                                                                           & \multicolumn{1}{c|}{\textbf{F1}} & \multicolumn{1}{c|}{\textbf{FAR}} & \multicolumn{1}{c|}{\textbf{FRR}} & \multicolumn{1}{c|}{\textbf{F1}} & \multicolumn{1}{c|}{\textbf{FAR}} & \multicolumn{1}{c|}{\textbf{FRR}} & \multicolumn{1}{c|}{\textbf{F1}} & \multicolumn{1}{c|}{\textbf{FAR}} & \multicolumn{1}{c|}{\textbf{FRR}} & \multicolumn{1}{c|}{\textbf{F1}} & \multicolumn{1}{c|}{\textbf{FAR}} & \multicolumn{1}{c|}{\textbf{FRR}} \\ \hline
\multicolumn{1}{|c|}{\multirow{3}{*}{\begin{tabular}[c]{@{}c@{}}Response \\ level\end{tabular}}} & Transcription & \multicolumn{1}{r|}{5.3}        & \multicolumn{1}{r|}{97.3}         & \multicolumn{1}{r|}{0}            & \multicolumn{1}{r|}{48.4}       & \multicolumn{1}{r|}{53.1}        & 46.9                             & \multicolumn{1}{r|}{76.1}       & \multicolumn{1}{r|}{27.0}        & \multicolumn{1}{r|}{20.6}        & \multicolumn{1}{r|}{83.1}       & \multicolumn{1}{r|}{8.6}         & 31.3                             \\ \cline{2-14} 
\multicolumn{1}{|c|}{}                                                                           & Paraphrasing  & \multicolumn{1}{r|}{56.3}       & \multicolumn{1}{r|}{52.1}        & \multicolumn{1}{r|}{22.4}         & \multicolumn{1}{r|}{64.8}        & \multicolumn{1}{r|}{38.6}        & 27.9                             & \multicolumn{1}{r|}{41.1}       & \multicolumn{1}{r|}{71.9}        & \multicolumn{1}{r|}{9.0}         & \multicolumn{1}{r|}{55.5}       & \multicolumn{1}{r|}{55.6}        & 16.1                             \\ \cline{2-14} 
\multicolumn{1}{|c|}{}                                                                           & Combined      & \multicolumn{1}{r|}{33.8}       & \multicolumn{1}{r|}{77.7}        & \multicolumn{1}{r|}{9.7}         & \multicolumn{1}{r|}{48.4}       & \multicolumn{1}{r|}{61.0}        & 22.6                              & \multicolumn{1}{r|}{50.5}       & \multicolumn{1}{r|}{62.6}        & \multicolumn{1}{r|}{11.2}        & \multicolumn{1}{r|}{55.8}       & \multicolumn{1}{r|}{56.4}        & 12.6                             \\ \hline
\multicolumn{1}{|c|}{\multirow{3}{*}{\begin{tabular}[c]{@{}c@{}}Window \\ level\end{tabular}}}   & Transcription & \multicolumn{1}{r|}{66.0}       & \multicolumn{1}{r|}{22.8}        & \multicolumn{1}{r|}{74.5}        & \multicolumn{1}{r|}{76.9}       & \multicolumn{1}{r|}{15.9}        & 46.2                             & \multicolumn{1}{r|}{69.2}       & \multicolumn{1}{r|}{20.1}        & \multicolumn{1}{r|}{66.8}        & \multicolumn{1}{r|}{77.3}       & \multicolumn{1}{r|}{25.0}        & 24.9                             \\ \cline{2-14} 
\multicolumn{1}{|c|}{}                                                                           & Paraphrasing  & \multicolumn{1}{r|}{7.5}         & \multicolumn{1}{r|}{96.1}        & \multicolumn{1}{r|}{1.4}         & \multicolumn{1}{r|}{21.8}       & \multicolumn{1}{r|}{87.3}        & 5.2                              & \multicolumn{1}{r|}{45}          & \multicolumn{1}{r|}{68.4}        & \multicolumn{1}{r|}{11.0}        & \multicolumn{1}{r|}{72.5}       & \multicolumn{1}{r|}{32.8}        & 22.5                             \\ \cline{2-14} 
\multicolumn{1}{|c|}{}                                                                           & Combined      & \multicolumn{1}{r|}{50.8}       & \multicolumn{1}{r|}{56.3}        & \multicolumn{1}{r|}{35.5}        & \multicolumn{1}{r|}{44.0}       & \multicolumn{1}{r|}{68.0}        & 16.7                             & \multicolumn{1}{r|}{54.5}       & \multicolumn{1}{r|}{55.9}        & \multicolumn{1}{r|}{22.6}        & \multicolumn{1}{r|}{68.9}       & \multicolumn{1}{r|}{31.9}        & 37.4                             \\ \hline
\end{tabular}
\end{table*}

\begin{table*}[htp]
\caption{Performance comparison of DetectGPT at window-level and sentence-level across datasets and proposed methods. Response-level evaluation is not reported because approximately half of the responses exceed the maximum context length supported by DetectGPT. Reported metrics include $F_1$ score, Accuracy, False Acceptance Rate (FAR), and False Rejection Rate (FRR).}
\label{tab:Window-Sentence-PerfDetectGPT}
\centering
\begin{tabular}{|c|r|r|r|r|r|r|r|r|}
\hline
\multirow{2}{*}{\textbf{Dataset / Method}} & \multicolumn{4}{c|}{\textbf{Window-level}} & \multicolumn{4}{c|}{\textbf{Sentence-level}} \\ \cline{2-9} 
 & \multicolumn{1}{c|}{\textbf{F1 (\%)}} & \multicolumn{1}{c|}{\textbf{Accuracy (\%)}} & \multicolumn{1}{c|}{\textbf{FAR (\%)}} & \multicolumn{1}{c|}{\textbf{FRR (\%)}} & \multicolumn{1}{c|}{\textbf{F1 (\%)}} & \multicolumn{1}{c|}{\textbf{Accuracy (\%)}} & \multicolumn{1}{c|}{\textbf{FAR (\%)}} & \multicolumn{1}{c|}{\textbf{FRR (\%)}} \\ \hline
SBU(GM)        & 6.0  & 58.3 & 29.7 & 51.3 & 63.4 & 53.2 & 63.5 & --    \\ \hline
SBU(GC)         & 52.0 & 59.4 & 48.7 & 34.5 & 53.2 & 59.9 & 74.5 & 19.1 \\ \hline
SBU(RF)          & 56.6 & 58.9 & 41.0 & 41.2 & 35.6 & 52.6 & 20.9 & 73.8 \\ \hline
IIITD-BU(Transcribed)        & 7.5  & 55.6 & 96.0 & 3.6  & 60.5 & 43.3 & 42.0 & 54.9 \\ \hline
IIITD-BU(Paraphrased)        & 10.9 & 49.0 & 94.0 & 5.0  & 41.1 & 53.4 & 32.3 & 63.8 \\ \hline
IIITD-BU(Transcribed + Paraphrased)    & 8.6  & 52.1 & 95.3 & 4.0  & 35.2 & 53.9 & 25.7 & 71.8 \\ \hline
\end{tabular}
\end{table*}

\subsubsection{Condition-Agnostic Performance}
We evaluate the condition-agnostic performance of ML-based detectors (LGBM and CatBoost) and the DL-based TypeNet across users, keyboards, contexts, and datasets. The results highlight clear differences in robustness and generalization across models.

\textit{User-agnostic scenarios.}
In user-agnostic evaluations, ML-based detectors achieve consistently high performance, with $F_1$ scores ranging from $97.0\%$ to $97.2\%$, FRR below $2.3\%$, and FAR close to $3.4\%$. These results indicate strong generalization across previously unseen users. In contrast, TypeNet attains a maximum $F_1$ score of $68.5\%$, with FAR and FRR values of $40.8\%$ and $43.8\%$, respectively, indicating limited robustness in user-independent settings.

\textit{Keyboard-agnostic scenarios.}
In keyboard-agnostic evaluations using the Buffalo dataset, ML-based detectors maintain strong performance, achieving $F_1$ scores between $98.0\%$ and $98.5$, with FRR consistently below $2.5\%$. TypeNet shows moderate improvement relative to the user-agnostic setting, achieving $F_1$ scores between $75.9\%$ and $78.8\%$, but remains substantially below the ML-based detectors. These results indicate reduced generalization of TypeNet when evaluated on previously unseen keyboards.

\textit{Context-agnostic scenarios.}
For context-agnostic evaluations conducted on the SBU dataset, ML-based detectors achieve moderate performance, with $F_1$ scores ranging from $80.2\%$ to $82.4\%$, FRR below $16\%$, and FAR near $22\%$. TypeNet achieves $F_1$ scores between $67.0\%$ and $76.3\%$. While ML-based detectors retain reasonable robustness across unseen contexts, TypeNet exhibits greater variability, suggesting reduced stability under cross-context conditions.

\textit{Dataset-agnostic scenarios.}
Dataset-agnostic evaluations reveal substantial variability across all models. ML-based detectors exhibit mixed performance, with LGBM achieving $F_1$ scores between $45.5\%$ and $84.1\%$, and CatBoost achieving $F_1$ scores between $62.7\%$ and $87.8\%$. TypeNet shows even greater variability, with $F_1$ scores ranging from $54.2\%$ to $71.1\%$. Performance differences are strongly influenced by dataset characteristics. Structured datasets such as SBU and Buffalo yield higher performance for ML-based detectors, whereas the IIITD-BU-Paraphrased dataset introduces linguistic and contextual variability that challenges all models. In dataset-agnostic evaluations, performance improves as the size of the training data increases, both in terms of user count and keystroke volume. Under this setting, ML-based detectors achieve a maximum $F_1$ score of $82.5\%$, while the proposed dataset yields the lowest $F_1$ score ($66.3\%)$, reflecting its comparatively smaller scale.

ML-based detectors, particularly CatBoost, exhibit strong performance across most evaluation scenarios. However, high FAR values indicate limited generalization, with the models tending to favor a single class during inference. This behavior is especially evident when the SBU and Buffalo datasets are used jointly for training or testing, while the IIITD-BU-Combined dataset is used in isolation or vice versa.

TypeNet shows marginal improvements in select scenarios but demonstrates lower overall performance and higher variability, reflecting its dependence on task-specific conditions and limited training data. The pronounced performance differences observed across datasets further highlight the influence of dataset characteristics. The SBU and Buffalo datasets, which are structured and do not include LLM-assisted plagiarism, are more amenable to ML-based detectors. In contrast, the IIITD-BU datasets introduce linguistically nuanced and LLM-assisted content that challenges both ML- and DL-based approaches. These findings indicate the need for improved robustness across heterogeneous datasets and deployment conditions.

\begin{figure*}[htp]
    \centering
    \includegraphics[width=6.7in, height = 1.66in]{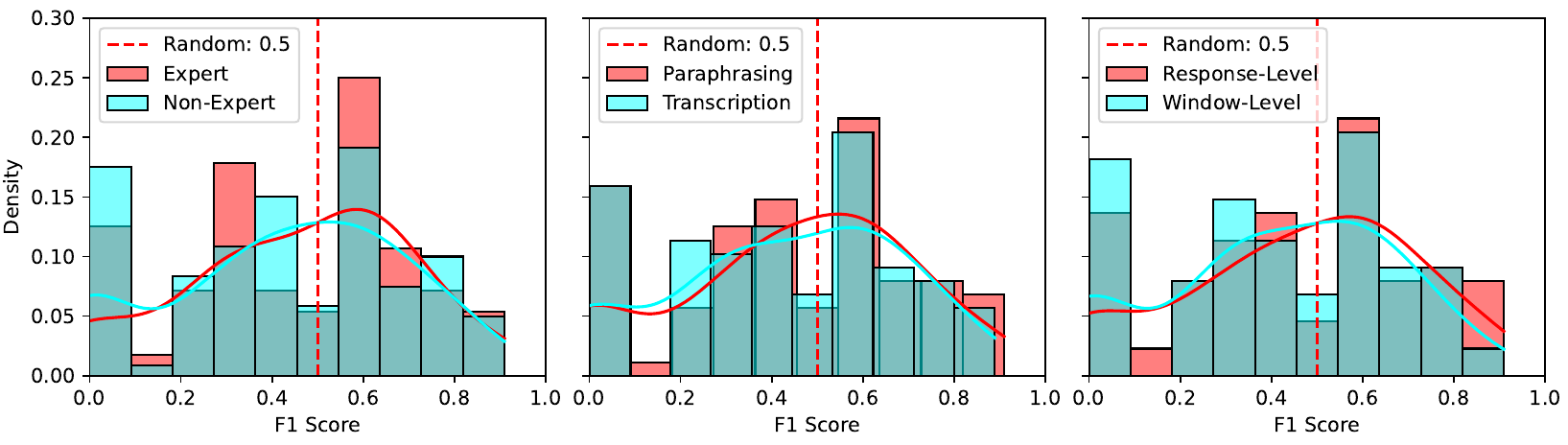}  
\caption{Histogram of $F_1$ scores illustrating the performance of human evaluators in plagiarism detection. From left to right: (a) comparison between expert and non-expert evaluators, (b) comparison between plagiarism styles (paraphrasing versus transcription), and (c) comparison between evaluation granularities (response-level versus window-level).}
    \label{fig:human-baseline}
\end{figure*}

\subsection{Performance Under Deception Scenarios}
The results show that the proposed deception-based attack is effective against ML-based detectors. As reported in Table~\ref{tab:attack}, both LGBM and CatBoost exhibit substantial vulnerability, with FRR exceeding $97\%$ under the pooled attack (At-P) and exceeding $93\%$ under the user-specific attack (At-U). In contrast, TypeNet attains $0\%$ FRR under both attack settings, despite having the highest baseline FRR ($29.2\%$) among all models prior to adversarial training.

\begin{figure}[h!]
  \centering
\includegraphics[width=3.25in, height=2.4in]{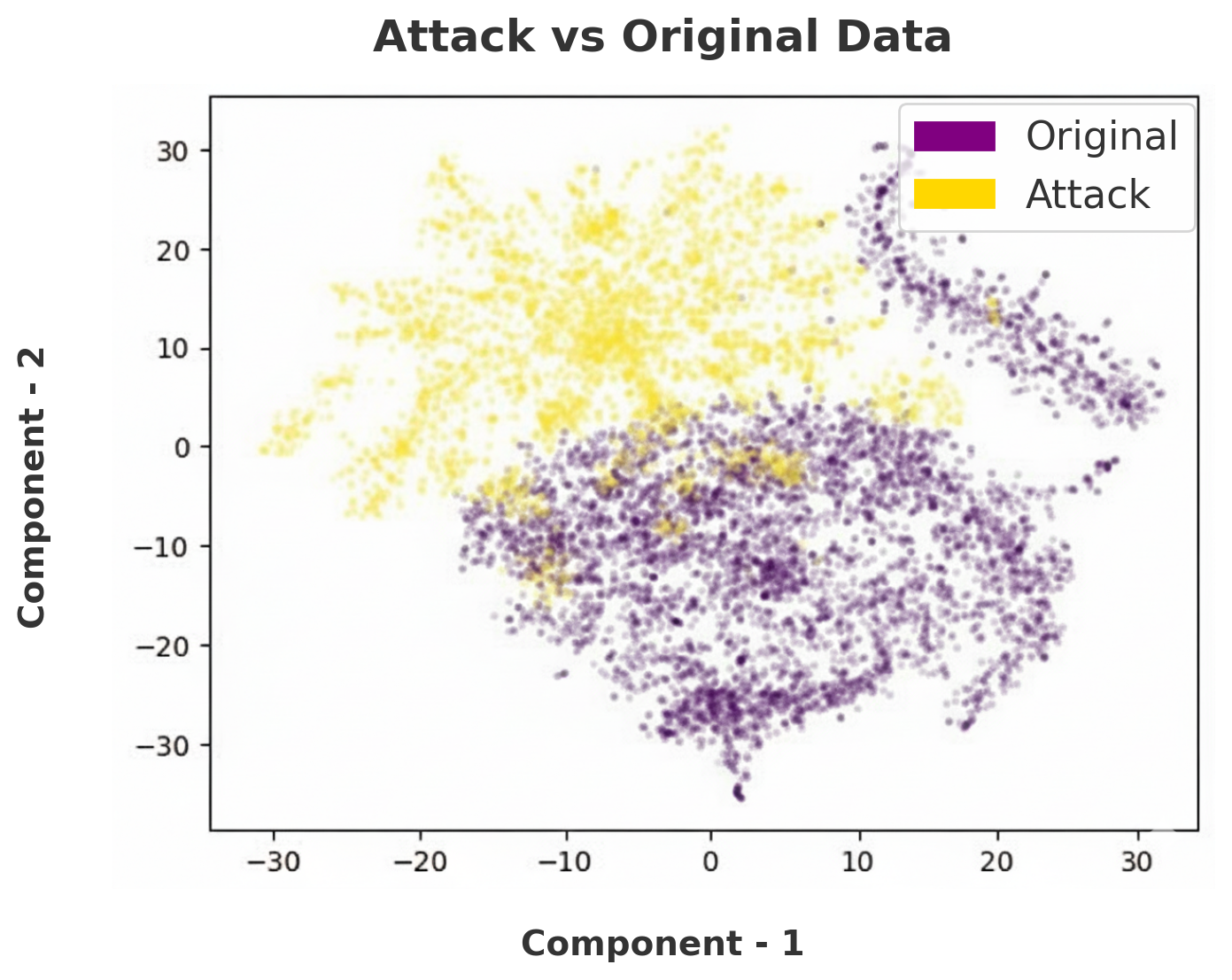}
  \caption{TSNE plot for the attack dataset with original data IIIT-BU-Paraphrased.}
  \label{fig:attack}
\end{figure}

\begin{figure}[h!]
  \centering
\includegraphics[width=3.3in, height=1.4in]{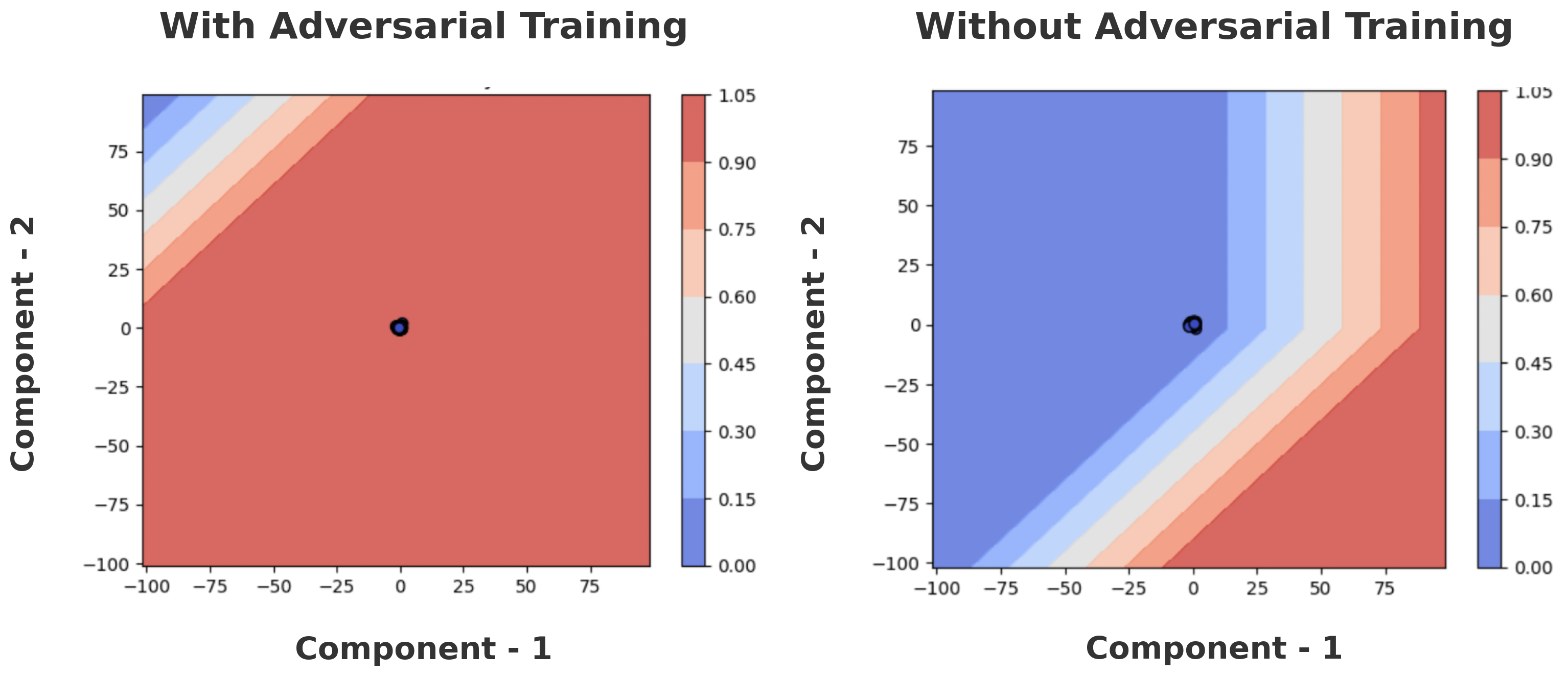}
  \caption{Decision boundaries: LGBM model decision boundaries before and after adversarial training. We sample and plot 50 attack samples after scaling and observe the changes in the decision boundary induced by adversarial training.}
\label{fig:decision_boundary}
\end{figure}

After adversarial training, the forged attack samples become highly separable, as illustrated in Figure~\ref{fig:attack}, enabling ML-based detectors to correctly classify all attack instances. To further examine this behavior, we visualize $50$ sampled data points and the corresponding decision boundaries before and after adversarial training (Figure~\ref{fig:decision_boundary}). Following training, the decision boundaries of the ML-based detectors shift to accommodate the forged samples, indicating direct adaptation to the attack patterns.

In contrast, TypeNet preserves a more stable feature representation learned through temporal sequence modeling, enabling it to generalize beyond the surface-level timing perturbations introduced by the attack. As a result, its decision boundary remains largely unchanged and continues to effectively separate bona fide and forged samples.

It is important to note that the observed robustness is specific to the modeled deception generator $\mathcal{G}$ and the associated At-U and At-P attack settings. Robustness to more adaptive or fundamentally different adversarial strategies is not implied.

\subsubsection{Performance of DetectGPT}
DetectGPT demonstrates limited effectiveness across the evaluated datasets, particularly under window-level evaluation (Table~\ref{tab:Window-Sentence-PerfDetectGPT}). On the IIITD-BU-Transcribed dataset, it achieves an $F_1$ score of $7.5\%$ with a FAR of $96.0\%$, and on the IIITD-BU-Paraphrased dataset, an $F_1$ score of $10.9\%$ with a FAR of $94.0\%$. Performance on the SBU dataset is similarly low under window-level evaluation, with an $F_1$ score of $6.0\%$.

Sentence-level evaluation yields higher $F_1$ scores, reaching $63.4\%$ on the SBU dataset and $60.5\%$ on the IIITD-BU-Transcribed dataset. However, these improvements are accompanied by substantially increased FRR, including $54.9\%$ on the IIITD-BU-Transcribed dataset and $63.8\%$ on the IIITD-BU-Paraphrased dataset. Overall, DetectGPT exhibits high rates of missed plagiarism under window-level evaluation and frequent false rejections of bona fide responses under sentence-level evaluation, indicating limited reliability of text-intrinsic detection in the presence of LLM-assisted plagiarism.

\subsection{Performance of LLaMA-3.3-70B-Instruct}
\label{LLaMAResults}
Table~\ref{tab:LLaMA-Instruct-Perf} reports the $F_1$ score, FRR, and FAR obtained by LLaMA-3.3-70B-Instruct under different learning paradigms, evaluation granularities, and input preparation settings. While preprocessing improves performance in several scenarios, the results indicate limitations that affect the model’s suitability for reliable plagiarism detection.

For transcription tasks at the response level, input preprocessing yields substantial performance gains. Under zero-shot learning, the $F_1$ score increases from $5.3\%$ to $76.1\%$, accompanied by a reduction in FRR from $97.3\%$ to $27.0\%$. Under few-shot learning, preprocessing increases the $F_1$ score from $48.4\%$ to $83.1\%$, while FRR decreases from $53.1\%$ to $8.6\%$.

For paraphrasing tasks, performance varies across evaluation granularities. At the response level, preprocessing degrades performance in the zero-shot setting, with the $F_1$ score decreasing from $56.3\%$ to $41.1\%$. Similar degradation is observed under few-shot learning, where the $F_1$ score decreases from $64.8\%$ to $55.5\%$. In contrast, preprocessing substantially improves performance at the window level. Zero-shot $F_1$ scores increase from $7.5\%$ to $45.0\%$, and few-shot $F_1$ scores increase from $21.8\%$ to $72.5\%$.

When transcription and paraphrasing samples are combined, preprocessing consistently improves performance across both evaluation granularities. At the response level, zero-shot $F_1$ increases from $33.8\%$ to $50.5\%$, and few-shot $F_1$ increases from $48.4\%$ to $55.8\%$. At the window level, zero-shot $F_1$ increases from $50.8\%$ to $54.5\%$, and few-shot $F_1$ increases from $44.0\%$ to $68.9\%$.

Despite these improvements, LLaMA-3.3-70B-Instruct exhibits high FRR in zero-shot settings and reduced effectiveness in complex scenarios, such as response-level paraphrasing, limiting its reliability as a standalone plagiarism-detection approach. In comparison, the proposed ML- and DL-based detectors achieve higher $F_1$ scores and more balanced FAR/FRR trade-offs across multiple scenarios.

Overall, preprocessing and few-shot prompting improve LLaMA-3.3-70B-Instruct performance in several settings; however, additional refinement is required to achieve consistent performance across diverse plagiarism scenarios.

\subsection{Performance of Human Evaluators}
\label{HumanPerfEval}
The $F_1$ scores indicate that human evaluators exhibit limited ability to reliably detect plagiarism across the evaluated scenarios. Among the four settings, response-level paraphrasing yields the highest performance ($F_1 = 48.4\%$), while window-level paraphrasing yields the lowest ($F_1 = 40.4\%$). For transcription tasks, performance is comparable across granularities, with $F_1$ scores of $43.0\%$ for response-level transcription and $42.2\%$ for window-level transcription. The small difference between expert ($F_1 = 45.5\%$) and non-expert ($F_1 = 42.6\%$) evaluators suggests that domain expertise does not substantially improve plagiarism detection performance. Overall, human performance remains close to chance level, underscoring the inherent difficulty of identifying plagiarism, particularly in paraphrasing scenarios.

Consistent with these observations, no clear performance advantage is observed among human evaluators across any condition (Figure~\ref{fig:human-baseline}). To further assess group-level differences, we conduct statistical tests of significance. A comparison between expert and non-expert evaluators shows no significant difference in $F_1$ score, as indicated by both a t-test ($t = 0.718, p = 0.474$) and a Mann-Whitney U test ($U = 3563.5, p = 0.518$). Similarly, the style of plagiarism (transcription versus paraphrasing) does not produce a significant effect on performance, with t-test results of $t = -0.467, p = 0.641$ and Mann-Whitney U test results of $U = 3692.5, p = 0.595$. Task granularity (response-level versus window-level) also shows no statistically significant difference, with a t-test result of $t = 1.148, p = 0.252$ and a Mann-Whitney U test result of $U = 4265.0, p = 0.244$.

These results indicate consistent performance across evaluator expertise, plagiarism styles, and task granularity, and highlight the limitations of human judgment in reliably distinguishing plagiarized content from bona fide writing in this setting.

\begin{table}[h!]
\caption{False Reject Rates (FRR) under different attack scenarios. At-U: user-specific attack, At-P: population attack.}
\centering
\begin{tabular}{|l|c|c|c|c|c|}
\hline
\multirow{2}{*}{\textbf{Model}} & \multirow{2}{*}{\textbf{Previous FRR}} & 
\multicolumn{2}{c|}{\textbf{Attack FRR}} & 
\multicolumn{2}{c|}{\textbf{Retrained FRR}} \\ \cline{3-6}
 & & \textbf{At-U} & \textbf{At-P} & \textbf{At-U} & \textbf{At-P} \\ \hline
LGBM     & 13.5\% & 97.3\% & 97.2\% & 0\% & 0\% \\ \hline
CatBoost & 14.9\% & 93.3\% & 96.8\% & 0\% & 0\% \\ \hline
TypeNet  & 29.2\% & 0\% & 0\% & -- & -- \\ \hline
\end{tabular}
\label{tab:attack}
\end{table}

\subsection{Performance Comparison and Caution}
Comparisons between the performance of the pretrained LLM LLaMA-3.3-70B-Instruct, human evaluators, and keystroke-based models should be interpreted with caution. Pretrained LLMs and human evaluators operate exclusively on the final textual output and therefore lack access to the behavioral process underlying text generation. As a result, their decisions are constrained to surface-level linguistic and stylistic cues.

In contrast, the proposed ML- and DL-based detectors leverage keystroke dynamics, which capture process-level behavioral information associated with text production. This additional modality provides access to temporal, cognitive, and motor patterns that are not observable from text alone. Consequently, keystroke-based approaches offer complementary capabilities for identifying plagiarism that may not be detectable through text-intrinsic analysis. While LLMs and human evaluators can identify stylistic inconsistencies, their reliance on static text limits their ability to detect behavioral anomalies indicative of assisted or plagiarized writing.

\subsection{Limitations and Ethical Considerations}
Despite demonstrating the potential of keystroke dynamics for detecting LLM-assisted plagiarism, this study has several limitations. First, the experimental evaluation is based on a limited number of users and datasets collected in controlled environments, which may not fully represent the variability encountered in real-world educational settings. Second, the datasets may not capture the full diversity of typing behaviors across different populations, tasks, and contexts, including creative writing, time-constrained assessments, or programming-oriented activities. Third, elevated error rates pose ethical and practical concerns. In particular, false positives may result in unjustified accusations of academic misconduct, potentially undermining trust in automated detection systems \cite{UnfairAccusations}. Notably, TypeNet exhibits relatively high error rates (28.9\%--45.8\%) in data- and topic-agnostic scenarios, limiting its reliability in such settings and indicating the need for further investigation and refinement.

In addition, detector performance is influenced by external factors such as keyboard layout and individual typing habits. Addressing these sources of variability remains an open challenge. Future work will focus on developing more robust models that adapt to heterogeneous user behaviors and hardware conditions. We also plan to explore integrating additional behavioral signals from typing tasks, such as mouse movement patterns and typing acoustics, to further improve detection robustness.

\section{Conclusion and Future Work}
\label{sec:conclusion_future_work}
This work demonstrates that keystroke dynamics provides a principled and effective foundation for detecting LLM-assisted academic dishonesty. By modeling the \emph{process} of text production rather than relying solely on the final textual artifact, keystroke-based detectors capture behavioral, cognitive, and motor patterns that are difficult to replicate through transcription or paraphrasing. This process-level perspective enables more reliable discrimination between bona fide writing and AI-assisted content, particularly in scenarios where text-only methods struggle.

Building on our prior work, we systematically expanded the dataset by including $90$ additional participants and introducing a paraphrasing condition that reflects realistic generative-AI-assisted plagiarism. Extensive benchmarking shows that gradient-boosted classifiers (LightGBM and CatBoost) achieve strong performance in structured settings, with $F_1$ scores exceeding $97\%$, while the sequence-based TypeNet model exhibits greater robustness for nuanced paraphrasing tasks, achieving an $F_1$ score of $86.9\%$. In contrast, text-intrinsic detectors, a 70B-parameter LLaMA-3.3-Instruct model, and human evaluators perform at or near chance in several settings, highlighting the limitations of relying solely on textual or subjective cues for authorship verification.

We further introduce a deception-based threat model that simulates forged keystroke traces and use it to analyze detector robustness. While ML-based detectors are highly vulnerable to such attacks, adversarial training substantially improves resilience, eliminating false rejections under the modeled threat. TypeNet demonstrates inherent robustness to these attacks because it relies on temporal sequence modeling. These results underscore both the promise of keystroke dynamics and the importance of explicitly evaluating robustness under adversarial conditions.

Future work will focus on improving generalizability by expanding datasets across more diverse populations, devices, and writing contexts. We also plan to integrate complementary behavioral signals, such as typing acoustics, mouse dynamics, eye gaze, and contextual metadata, to further strengthen detection. Finally, exploring hybrid models that combine linguistic and behavioral signals, alongside privacy-preserving feature extraction and explainable decision-making mechanisms, will be critical for deploying fair, transparent, and ethically responsible academic integrity systems in environments increasingly shaped by generative AI.

\ifCLASSOPTIONcompsoc
  \section*{Acknowledgments}
\else
  \section*{Acknowledgment}
\fi
We sincerely thank the anonymous IJCB and TBIOM reviewers for their valuable and constructive feedback. Their insights have significantly improved the quality and clarity of our manuscript.
 
\balance
\bibliography{short_references.bib}

@misc{geminiteam2025,
      title={Gemini: A Family of Highly Capable Multimodal Models}, 
      author={Gemini Team},
      year={2025},
      eprint={2312.11805},
      archivePrefix={arXiv},
      primaryClass={cs.CL},
      url={https://arxiv.org/abs/2312.11805}, 
}

@inproceedings{jemma2025how,
  author    = {Rebira Jemma and Rajesh Kumar},
  title     = {How Well Do LLMs Imitate Human Writing Style?},
  booktitle = {IEEE -UEMCON},
  year      = {2025},
  address   = {New York, NY, United States}
}

@inproceedings{roh2025llm,
  author    = {D. Roh and R. Kumar and A. Ngo},
  title     = {LLM-Assisted Cheating Detection in Korean Language via Keystrokes},
  booktitle = {IEEE IJCB},
  year      = {2025},

}

@inproceedings{khan2018augmented,
  title={Augmented reality-based mimicry attacks on behaviour-based smartphone authentication},
  author={Khan, Hassan and Hengartner, Urs and Vogel, Daniel},
  booktitle={ACM MobiSys},
  year={2018}
}

@article{MimicryAttacksKeystrokes,
author = {Khan, Hassan and Hengartner, Urs and Vogel, Daniel},
year = {2020},
title = {Mimicry Attacks on Smartphone Keystroke Authentication},
 
journal = {ACM TOPS},
 
}

@ARTICLE{ChatGPTOnEducation,
AUTHOR={Dempere, Juan  and Modugu, Kennedy  and Hesham, Allam  and Ramasamy, Lakshmana Kumar },
TITLE={The impact of ChatGPT on higher education},
JOURNAL={Frontiers in Education},
YEAR={2023},
 }

@article{trezise2019contract_cheating_detecting,
  title={Detecting Contract Cheating Using Learning Analytics},
  author={Trezise, Kelly and Ryan, Tracii and de Barba, Paula and Kennedy, Gregor},
  journal={Journal of Learning Analytics},
  year={2019},
  publisher={Society for Learning Analytics Research},
}

@INPROCEEDINGS{agarwal-nancy-contract,
  author={Agarwal, Nancy and Danielsen, Nils Folvik and Gravdal, Per Kristian and Bours, Patrick},
  booktitle={ICETA}, 
  title={Contract Cheat Detection using Biometric Keystroke Dynamics}, 
  year={2022},
}

@misc{Writefull,
  author = {{Writefull}},
  title  = {Writefull},
  year   = {2024},
  howpublished    = {\url{https://www.writefull.com/}},
}

@misc{plurilockKeystrokeDynamicsDiffMimic,
	author = {Plurilock},
	title = {{K}eystroke {D}ynamics - {P}lurilock --- plurilock.com},
	howpublished = {\url{https://plurilock.com/deep-dive/keystroke-dynamics/}},
	year = {2024},
	note = {[Accessed 31-12-2024]},
}

@book{adelani2020generating,
	author = {Adelani, David Ifeoluwa and Mai, Haotian and Fang, Fuming and Nguyen, Huy H. and Yamagishi, Junichi and Echizen, Isao},
	booktitle = {Advances in intelligent systems and computing},
	title = {{Generating Sentiment-Preserving fake online reviews using neural language models and their human- and Machine-Based detection}},
	year = {2020},
 
}

@inproceedings{tan2014effect,
    title = "The effect of wording on message propagation: Topic- and author-controlled natural experiments on {T}witter",
    author = "Tan, Chenhao  and
      Lee, Lillian  and
      Pang, Bo",
    editor = "Toutanova, Kristina  and
      Wu, Hua",
    month = jun,
    year = "2014",
    address = "Baltimore, Maryland",
    publisher = "ACL",
 
}

@incollection{tetlock2017expert,
	title        = {Expert political judgment},
	author       = {Tetlock, Philip E},
	year         = 2017,
	booktitle    = {Expert Political Judgment},
	publisher    = {Princeton University Press}
}

@online{hindu2024llm,
  author = {The Hindu Bureau},
  title = {LLM student sues Jindal Global Law School over AI use in examination},
  year = {2024},
  howpublished = {\url{https://shorturl.at/vP2fh}},
  note = {[Accessed 25-01-2026] }

}

@article{hayes1981uncovering,
	author = {Hayes, John R. and Flower, Linda},
	journal = {ERIC Clearinghouse},
	title = {{Uncovering Cognitive Processes in Writing: An Introduction to Protocol Analysis.}},
	year = {1981},
	url = {https://eric.ed.gov/?id=ED202035},
}

@inproceedings{singh2024llava,
  title={Teaching Human Behavior Improves Content Understanding Abilities Of VLMs},
  author={Singh, Somesh Kumar and Singla, Yaman Kumar and Chen, Changyou and Shah, Rajiv Ratn and Baths, Veeky and Krishnamurthy, Balaji and others},
  booktitle={ICLR},
  year={2025}
}

@article{sood2020improving,
  title={Improving natural language processing tasks with human gaze-guided neural attention},
  author={Sood, Ekta and Tannert, Simon and M{\"u}ller, Philipp and Bulling, Andreas},
  journal={NeurIPS'20},
 
}

@inproceedings{khurana2023synthesizing,
  title={Synthesizing human gaze feedback for improved NLP performance},
  author={Khurana, Varun and Kumar, Yaman and Hollenstein, Nora and Kumar, Rajesh and Krishnamurthy, Balaji},
  booktitle={EACL},
  pages={1895--1908},
  year={2023}
}

@inproceedings{plank2016keystroke,
  title={Keystroke dynamics as signal for shallow syntactic parsing},
  author={Plank, Barbara},
  booktitle={COLING},
  year={2016},
  organization={ACL}
}

@online{tenbarge2024ai,
  author = {Kat Tenbarge},
  title = {Parents sue son’s high school history teacher over AI ‘cheating’ punishment},
  year = {2024},
  month = {October},
  howpublished    = {\url{https://shorturl.at/6ebwO}},
}

@online{coley2023guidance,
  author = {Michael Coley},
  title = {Guidance on AI Detection and Why We’re Disabling Turnitin’s AI Detector},
  year = {2023},
  month = {August},
  howpublished    = {\url{https://shorturl.at/rCe4x}}
}

@online{young2024grammarly,
  author = {Jeffrey R. Young},
  title = {What happened after this college student’s paper was falsely flagged for AI use after using Grammarly},
  year = {2024},
  month = {February},
  howpublished    = {\url{https://shorturl.at/jt3HC}},
  note = {[Accessed 25-01-2026] }

}

@online{coffey2024professors,
  author = {Lauren Coffey},
  title = {Professors Cautious of Tools to Detect AI-Generated Writing},
  year = {2024},
  month = {February},
  day = {9},
  howpublished    = {\url{https://shorturl.at/PBvTf}},
  note = {[Accessed 25-01-2026] }

}

@article{isola2013makes,
	title        = {What makes a photograph memorable?},
	author       = {Isola, Phillip and Xiao, Jianxiong and Parikh, Devi and Torralba, Antonio and Oliva, Aude},
	year         = 2013,
	journal      = {IEEE TPAMI},
}

@article{foltynek2019academic,
  title={Academic plagiarism detection: a systematic literature review},
  author={Folt{\`y}nek, Tom{\'a}{\v{s}} and Meuschke, Norman and Gipp, Bela},
  journal={ACM CSUR},
  year={2019},
  publisher={ACM New York, NY, USA}
}

@article{weiss2019deepfake,
  title={Deepfake bot submissions to federal public comment websites cannot be distinguished from human submissions},
  author={Weiss, Max},
  journal={Technology Science},
  year={2019}
}

@article{lavergne2008detecting,
  title={Detecting Fake Content with Relative Entropy Scoring.},
  author={Lavergne, Thomas and Urvoy, Tanguy and Yvon, Fran{\c{c}}ois},
  journal={Pan},
  year={2008}
}

@article{crawford2015survey,
  title={Survey of review spam detection using machine learning techniques},
  author={Crawford, Michael and Khoshgoftaar, Taghi M and Prusa, Joseph D and Richter, Aaron N and Al Najada, Hamzah},
  journal={Journal of Big Data},
  year={2015},
  publisher={Springer}
}

@article{mridha2021comprehensive,
  title={A comprehensive review on fake news detection with deep learning},
  author={Mridha, Muhammad Firoz and Keya, Ashfia Jannat and Hamid, Md Abdul and Monowar, Muhammad Mostafa and Rahman, Md Saifur},
  journal={IEEE Access},
  year={2021},
}

@inproceedings{mitchell2023detectgpt,
  title={Detectgpt: Zero-shot machine-generated text detection using probability curvature},
  author={Mitchell, Eric and Lee, Yoonho and Khazatsky, Alexander and Manning, Christopher D and Finn, Chelsea},
  booktitle={ICML},
  year={2023},
}

@inproceedings{gehrmann2019gltr,
  title={GLTR: Statistical Detection and Visualization of Generated Text},
  author={Gehrmann, Sebastian and Strobelt, Hendrik and Rush, Alexander},
  booktitle={ACL: System Demonstrations},
  year={2019},
}

@article{solaiman2019release,
  title={Release strategies and the social impacts of language models},
  author={Solaiman, Irene and Brundage, Miles and Clark, Jack and Askell, Amanda and Herbert-Voss, Ariel and Wu, Jeff and Radford, Alec and Krueger, Gretchen and Kim, Jong Wook and Kreps, Sarah and others},
  journal={arXiv preprint arXiv:1908.09203},
  year={2019}
}

@inproceedings{si2023long,
  title={Long-Term Ad Memorability: Understanding \& Generating Memorable Ads},
  author={Harini, SI and Singh, Somesh and Singla, Yaman K and Bhattacharyya, Aanisha and Baths, Veeky and Chen, Changyou and Shah, Rajiv Ratn and Krishnamurthy, Balaji},
  booktitle={WACV},
  year={2025},
  organization={IEEE}
}

@article{matsuhashi1981pausing,
  title={Pausing and planning: The tempo of written discourse production},
  author={Matsuhashi, Ann},
  journal={Research in the Teaching of English},
  year={1981},
  publisher={ncte. org}
}

@article{van2008pause,
  title={Pause time patterns in writing narrative and expository texts by children and adults},
  author={Van Hell, Janet G and Verhoeven, Ludo and Van Beijsterveldt, Liesbeth M},
  journal={Discourse Processes},
  year={2008},
  publisher={Taylor \& Francis}
}

@article{ekman2003darwin,
  title={Darwin, deception, and facial expression},
  author={Ekman, Paul},
  journal={Annals of the New York Academy of Sciences},
  year={2003},
  publisher={Wiley Online Library}
}

@inproceedings{ott2011finding,
author = {Ott, Myle and Choi, Yejin and Cardie, Claire and Hancock, Jeffrey T.},
title = {Finding deceptive opinion spam by any stretch of the imagination},
year = {2011},
isbn = {9781932432879},
publisher = {Association for Computational Linguistics},
address = {USA},
booktitle = {ACL: Human Language Technologies - Volume 1},
series = {HLT '11}
}

@inproceedings{clark2021all,
  title={All that’s ‘Human’is not gold: Evaluating human evaluation of generated text.},
  author={Clark, E},
  booktitle={ACL-IJCNLP},
  volume={1},
  pages={7282},
  year={2021}
}

@inproceedings{ippolito2019automatic,
  title={Automatic detection of generated text is easiest when humans are fooled},
  author={Ippolito, Daphne and Duckworth, Daniel and Callison-Burch, Chris and Eck, Douglas},
  booktitle={ACL},
  year={2020}
}

@article{baaijen2012keystroke,
  title={Keystroke analysis: Reflections on procedures and measures},
  author={Baaijen, Veerle M and Galbraith, David and De Glopper, Kees},
  journal={Written Communication},
  year={2012},
  publisher={Sage Publications Sage CA: Los Angeles, CA}
}

@incollection{wengelin2006examining,
  title={Examining pauses in writing: Theory, methods and empirical data},
  author={Wengelin, {\AA}sa},
  booktitle={Computer key-stroke logging and writing},
  year={2006},
  publisher={Brill}
}

@article{van2009keystroke,
  title={Keystroke logging in writing research: Observing writing processes with Inputlog},
  author={Van Waes, Luuk and Leijten, Mari{\"e}lle and Van Weijen, Daphne},
  journal={German as a foreign language},
  year={2009}
}

@online{merod2023turnitin,
  author = {Anna Merod},
  title = {Turnitin admits there are some cases of higher false positives in AI writing detection tool},
  year = {2023},
  howpublished = {\url{https://www.k12dive.com/news/turnitin-false-positives-AI-detector/652221/}},
  note = {[Accessed 25-01-2026] }

}

@inproceedings{kirchenbauer2023watermark,
  title={A watermark for large language models},
  author={Kirchenbauer, John and Geiping, Jonas and Wen, Yuxin and Katz, Jonathan and Miers, Ian and Goldstein, Tom},
  booktitle={ICML},
  year={2023},
}

@inproceedings{atallah2001natural,
  title={Natural language watermarking: Design, analysis, and a proof-of-concept implementation},
  author={Atallah, Mikhail J and Raskin, Victor and Crogan, Michael and Hempelmann, Christian and Kerschbaum, Florian and Mohamed, Dina and Naik, Sanket},
  booktitle={Information Hiding: 4th International Workshop},
  year={2001},
  organization={Springer}
}

@inproceedings{singh2024measuring,
  title={Measuring And Improving Persuasiveness Of Large Language Models},
  author={Singh, Somesh Kumar and Singla, Yaman Kumar and Krishnamurthy, Balaji and others},
  booktitle={ICLR}
}

@PHDTHESIS{Sharma2023CNNKeystrokes,
  title     = "Keystroke Dynamics and User Identification",
  author    = "Sharma, Atharva",
  publisher = "San Jose State University Library",
  year      =  2023
}

@ARTICLE{UnfairAccusations,
  title     = "The problem with false positives: {AI} detection unfairly
               accuses scholars of {AI} plagiarism",
  author    = "Giray, Louie",
  journal   = "Ser. Libr.",
  publisher = "Informa UK Limited",
  month     =  dec,
  year      =  2024,
  language  = "en"
}

@inproceedings{CatBoost,
author = {Prokhorenkova, Liudmila and Gusev, Gleb and Vorobev, Aleksandr and Dorogush, Anna Veronika and Gulin, Andrey},
title = {CatBoost: unbiased boosting with categorical features},
year = {2018},
publisher = {Curran Associates Inc.},
booktitle = {NeurIPS},
}

@inproceedings{LGBM,
author = {Ke, Guolin and Meng, Qi and Finley, Thomas and Wang, Taifeng and Chen, Wei and Ma, Weidong and Ye, Qiwei and Liu, Tie-Yan},
title = {LightGBM: a highly efficient gradient boosting decision tree},
year = {2017},
booktitle = {NeurIPS},

}

@article{keystrokesCNN,
title = {Continuous authentication by free-text keystroke based on CNN and RNN},
journal = {Computers and Security},
year = {2020},
author = {Xiaofeng Lu and Shengfei Zhang and Pan Hui and Pietro Lio},
}

@book{anderson2001taxonomy,
  title     = {A Taxonomy for Learning, Teaching, and Assessing: A Revision of Bloom's Taxonomy of Educational Objectives},
  author    = {Anderson, Lorin W. and Krathwohl, David R. and Airasian, Peter W. and Cruikshank, Kathleen A. and Mayer, Richard E. and Pintrich, Paul R. and Raths, James and Wittrock, Merlin C.},
  year      = {2001},
  publisher = {Addison Wesley Longman, Inc.},
  address   = {New York},
}

@article{kundu2024keystroke,
  title={Keystroke Dynamics Against Academic Dishonesty in the Age of LLMs},
  author={Kundu, Debnath and Mehta, Atharva and Kumar, Rajesh and Lal, Naman and Anand, Avinash and Singh, Apoorv and Shah, Rajiv Ratn},
  journal={IEEE IJCB},
  year={2024},
}

@article{yan2023detectionEssayWriting,
  title={Detection of AI-generated Essays in Writing Assessments},
  author={Yan, Duanli and Fauss, Michael and Hao, Jiangang and Cui, Wenju},
  journal={Psychological Test and Assessment Modeling},
  year={2023},
  publisher={Educational Testing Service}
}

@INPROCEEDINGS{TypingPhoneBTAS2016,
  author={Kumar, Rajesh and Phoha, Vir V. and Serwadda, Abdul},
  booktitle={IEEE International Conference on Biometrics: Theory, Applications, and Systems}, 
  title={Continuous authentication of smartphone users by fusing typing, swiping, and phone movement patterns}, 
  year={2016},
}

@article{teh2013survey,
  title={A survey of keystroke dynamics biometrics},
  author={Teh, Pin Shen and Teoh, Andrew Beng Jin and Yue, Shigang and others},
  journal={The Scientific World Journal},
  year={2013},
  publisher={Hindawi}
}

@article{zellers2019defending,
  title={Defending against neural fake news},
  author={Zellers, Rowan and Holtzman, Ari and Rashkin, Hannah and Bisk, Yonatan and Farhadi, Ali and Roesner, Franziska and Choi, Yejin},
  journal={NeurIPS},
  year={2019}
}

@inproceedings{singla2022minimal,
  title={Minimal: Mining models for universal adversarial triggers},
  author={Singla, Yaman Kumar and Parekh, Swapnil and Singh, Somesh and Chen, Changyou and Krishnamurthy, Balaji and Shah, Rajiv Ratn},
  booktitle={AAAI},
  year={2022}
}

@article{kumar2023automatic,
  title={Automatic essay scoring systems are both overstable and oversensitive: explaining why and proposing defenses},
  author={Yaman Kumar et al.},
  journal={Dialogue \& Discourse},
  year={2023}
}

@article{frohling2021feature,
  title={Feature-based detection of automated language models: tackling GPT-2, GPT-3 and Grover},
  author={Fr{\"o}hling, Leon and Zubiaga, Arkaitz},
  journal={PeerJ Computer Science},
  year={2021},
  publisher={PeerJ Inc.}
}

@inproceedings{uchendu2020authorship,
  title={Authorship attribution for neural text generation},
  author={Uchendu, Adaku and Le, Thai and Shu, Kai and Lee, Dongwon},
  booktitle={EMNLP},
  year={2020}
}

@article{fagni2021tweepfake,
  title={TweepFake: About detecting deepfake tweets},
  author={Fagni, Tiziano and Falchi, Fabrizio and Gambini, Margherita and Martella, Antonio and Tesconi, Maurizio},
  journal={Plos one},
  year={2021},
  publisher={Public Library of Science San Francisco, CA USA}
}

@inproceedings{dugan2023real,
  title={Real or fake text?: Investigating human ability to detect boundaries between human-written and machine-generated text},
  author={Dugan, Liam and Ippolito, Daphne and Kirubarajan, Arun and Shi, Sherry and Callison-Burch, Chris},
  booktitle={AAAI},
  year={2023}
}

@inproceedings{kuruvilla2024spotting,
  title={Spotting Fake Profiles in Social Networks via Keystroke Dynamics},
  author={Kuruvilla, Alvin and Daley, Rojanaye and Kumar, Rajesh},
  booktitle={IEEE-CCNC},
  year={2024},
}

@article{acien2022detection,
  title={Detection of mental fatigue in the general population: Feasibility study of keystroke dynamics as a real-world biomarker},
  author={Acien, Alejandro et al.},
  journal={JMIR Biomedical Engineering},
  year={2022},
  publisher={JMIR Publications Inc., Toronto, Canada}
}

@article{vizer2009automated,
  title={Automated stress detection using keystroke and linguistic features: An exploratory study},
  author={Vizer, Lisa M and Zhou, Lina and Sears, Andrew},
  journal={International Journal of Human-Computer Studies},
  year={2009},
  publisher={Elsevier}
}

@InProceedings{banerjee2014_emnlp,
	author    = {Banerjee, Ritwik  and  Feng, Song  and  Kang, Jun Seok  and  Choi, Yejin},
	title     = {Keystroke Patterns as Prosody in Digital Writings: A Case Study with Deceptive Reviews and Essays},
	booktitle = {EMNLP},
 	year      = {2014},
	address   = {Doha, Qatar},
 }

@INPROCEEDINGS{7823894,
  author={Sun, Yan and Ceker, Hayreddin and Upadhyaya, Shambhu},
  booktitle={WIFS}, 
  title={Shared keystroke dataset for continuous authentication}, 
  year={2016},
}

@article{shadman2023keystroke,
  title={Keystroke dynamics: Concepts, techniques, and applications},
  author={Shadman, Rashik and Wahab, Ahmed Anu and Manno, Michael and Lukaszewski, Matthew and Hou, Daqing and Hussain, Faraz},
  journal={ACM Computing Surveys},
  year={2025},
}

@ARTICLE{acien2021typenet,
  author={Acien, Alejandro and Morales, Aythami and Monaco, John V. and Vera-Rodriguez, Ruben and Fierrez, Julian},
  journal={The IEEE TBIOM}, 
  title={TypeNet: Deep Learning Keystroke Biometrics}, 
  year={2022},
}

@inproceedings{wahab2023simple,
  title={When Simple Statistical Algorithms Outperform Deep Learning: A Case of Keystroke Dynamics},
  author={Wahab, Ahmed and Hou, Daqing},
  booktitle={The ICPRAM},
  year={2023}
}

@ARTICLE{TPAMIKeystroke1990,
  author={Bleha, S. and Slivinsky, C. and Hussien, B.},
  journal={IEEE TPAMI}, 
  title={Computer-access security systems using keystroke dynamics}, 
  year={1990},
}

@article{press1980authentication,
  title={Authentication by keystroke timing: Some preliminary results},
  author={Press, S},
  journal={Rand Report},
  year={1980}
}

@inproceedings{itad_ref0,
  author    = {Blaine Ayotte and Mahesh K. Banavar and Daqing Hou and Stephanie Schuckers},
  title     = {Fast and Accurate Continuous User Authentication by Fusion of Instance-based, Free-text Keystroke Dynamics},
  booktitle = {Proc. BIOSIG 2019 -- Int. Conf. Biometrics Special Interest Group},
  series    = {Lecture Notes in Informatics (LNI)},
  editor    = {A. Br{\"o}mme and C. Busch and A. Dantcheva and C. Rathgeb and A. Uhl},
  address   = {Bonn, Germany},
  year      = {2019},
}

@ARTICLE{itad_ref,
  author={Ayotte, Blaine and Banavar, Mahesh and Hou, Daqing and Schuckers, Stephanie},
  journal={The IEEE TBIOM}, 
  title={Fast Free-Text Authentication via Instance-Based Keystroke Dynamics}, 
  year={2020},}

@article{GONZALEZ2023100454,
title = {KSDSLD — A tool for keystroke dynamics synthesis \
         \& liveness detection},
journal = {Software Impacts},
year = {2023},
author = {Nahuel González},
}

@article{Gunetti2005KeystrokeAO,
  title={Keystroke analysis of free text},
  author={Daniele Gunetti and Claudia Picardi},
  journal={ACM TISSEC},
  year={2005},
  url={https://api.semanticscholar.org/CorpusID:13904989}
}

@article{phoha2010methods,
  title={Methods of identifying users based on text entered on keyboard},
  author={Phoha, V and Joshi, S},
  journal={Patent Pending' 10},
 
}

@misc{gdb,
  author = {{OnlineGDB}},
  title  = {OnlineGDB: Online Compiler and Debugger},
  howpublished    = {\url{https://www.onlinegdb.com/}},
  note = {[Accessed 25-01-2026] }

}

@software{ChatGPT,
  author       = {{OpenAI}},
  title        = {ChatGPT},
  year         = {2024},
  version      = {GPT-4o},
  howpublished          = {https://chat.openai.com/},
  note = {[Accessed 25-01-2026] }

}

@software{Claude,
  author       = {{Anthropic}},
  title        = {Claude 3 Opus},
  year         = {2024},
  howpublished          = {https://www.anthropic.com/news/claude-3-family},
  note         = {[Accessed 25-01-2026] }

}

@article{grattafiori2024llama,
  title={The llama 3 herd of models},
  author={Grattafiori, Aaron et al.},
  journal={arXiv preprint arXiv:2407.21783},
  year={2024}
}

@software{zerogpt,
  author = {{ZeroGPT}},
  title  = {ZeroGPT: AI Content Detection Tool},
  year   = {2024},
  howpublished    = {\url{https://www.zerogpt.com/}},
  note = {[Accessed 25-01-2026] }

}

@software{quill,
  author = {{QuillBot}},
  title  = {QuillBot AI Content Detector},
  year   = {2024},
  howpublished    = {\url{https://quillbot.com/ai-content-detector}},
  note = {[Accessed 25-01-2026] }
}

@software{gptzero,
  author = {{GPTZero}},
  title  = {GPTZero: AI Text Detection Tool},
  year   = {2024},
  howpublished    = {\url{https://gptzero.me/}},
  note = {[Accessed 25-01-2026] }

}

@article{Canyakan_2025, title={Comparative accuracy of AI-based plagiarism detection tools: an enhanced systematic review}, volume={1}, url={https://jaihne.com/index.php/jaihne/article/view/11}, author={Canyakan, Seyhan}, year={2025}, month={Mar.}, pages={5–18} }

@article{ai_academic_integrity,
author = {Leong, Wai Yie and Zhang, J},
year = {2025},
month = {03},
pages = {2025},
title = {AI on Academic Integrity and Plagiarism Detection},
volume = {20},
journal = {ASM Science Journal},
doi = {10.32802/asmscj.2025.1918}
}
\bibliographystyle{unsrt}
    \vspace{0.15in}

    \hangindent=\dimexpr 1.1in+1em\relax
    \hangafter=-10
    \noindent\llap{\raisebox{\dimexpr\ht\strutbox-\height}[0pt][0pt]{\includegraphics[width=1in,height=1in]{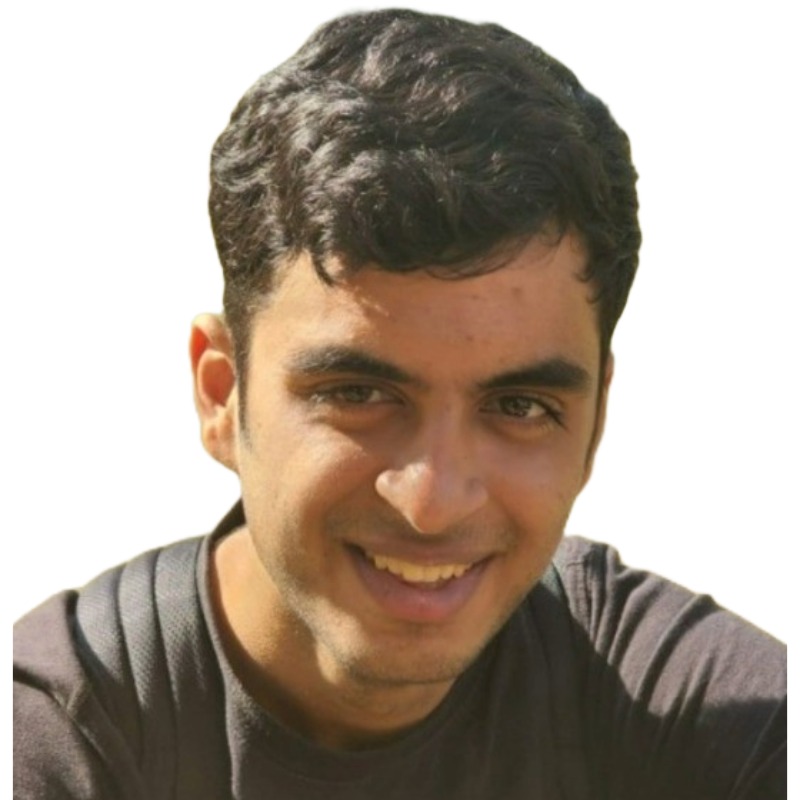}}
    \hspace{1em}}
    \scriptsize \textbf{Atharva Mehta} is a Research Engineer at MBZUAI, Abu Dhabi. He completed his Bachelor’s degree from IIIT Delhi. His work focuses on uncovering and addressing bias in music generation models, on ethics in music AI, on developing techniques for low-resource music generation, and on exploring creativity and aesthetics through computational approaches to music. Beyond music, his research extends to understanding social norms and human behavior through natural language processing, particularly via linguistic phenomena such as honorifics, and through biometric analysis of keystroke dynamics.
    
    \hangindent=\dimexpr 1.1in+1em\relax
    \hangafter=-10
    \noindent\llap{\raisebox{\dimexpr\ht\strutbox-\height}[0pt][0pt]{\includegraphics[width=1in,height=1in]{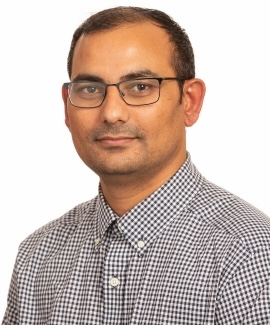}}%
    \hspace{1em}}
    \scriptsize \textbf{Dr. Rajesh Kumar} is an Assistant Professor of Computer Science at Bucknell University, USA, and an Adjunct Faculty at IIIT Delhi, India. He received the Ph.D. degree in Computer Science from Syracuse University, USA, the M.S. degree in Mathematics from Louisiana Tech University, USA, and the M.C.A. degree from Jawaharlal Nehru University, New Delhi, India. His research leverages everyday human activities such as walking, typing, and swiping to advance cybersecurity, social media forensics, and academic integrity. His work has appeared in leading IEEE and ACM conferences and transactions, and has received the Best Paper award, as well as multiple Best-Reviewed Paper and Best Poster awards. He has also been recognized with several Best Reviewer Awards. \href{https://sites.google.com/view/kumar7}{Click here for more details.}
    
    \hangindent=\dimexpr 1.1in+1em\relax
    \hangafter=-10
    \noindent\llap{\raisebox{\dimexpr\ht\strutbox-\height}[0pt][0pt]{\includegraphics[width=1in,height=1in]{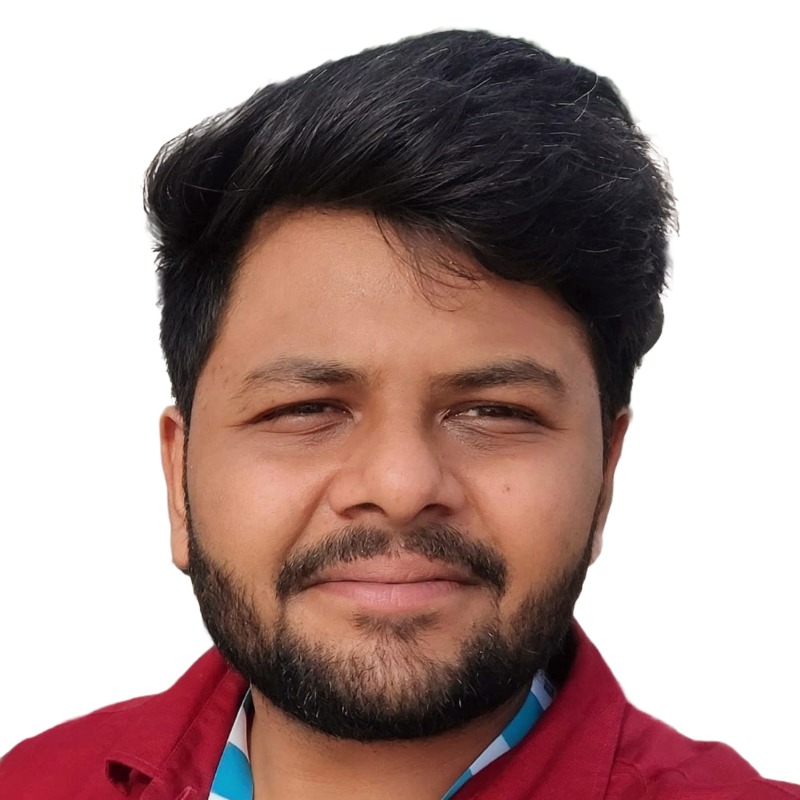}}%
    \hspace{1em}}
    \scriptsize \textbf{Aman Singla} is a Research Associate at the Indraprastha Institute of Information Technology, Delhi (IIIT-Delhi), India. He received his Bachelor’s degree in Computer Science from Maharishi Dayanand University, Haryana, India. His research interests lie at the intersection of technology, design, and data-driven systems, with a particular focus on developing scalable, intelligent solutions. He has contributed to projects such as Audino, where he built intelligent systems for data annotation and automation, streamlining large-scale audio and text processing workflows. Mr. Singla is a passionate technologist and problem solver, committed to continuous learning, collaboration, and creating products that bridge the gap between human creativity and technology.
        
    \hangindent=\dimexpr 1.1in+1em\relax
    \hangafter=-22
    \noindent\llap{\raisebox{\dimexpr\ht\strutbox-\height}[0pt][0pt]{\includegraphics[width=1in,height=1in]{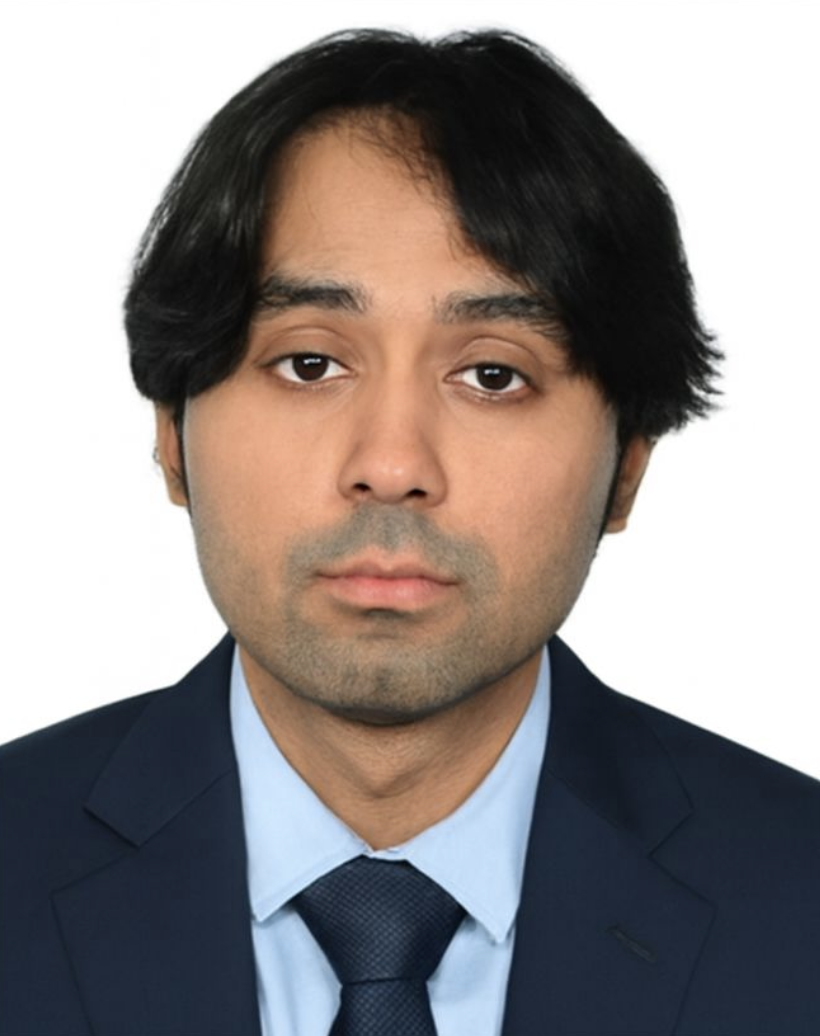}}%
    \hspace{1em}}
    \scriptsize \textbf{Karthik Bisht} is a Research Associate at the Multimodal Digital Media Analysis (MIDAS) Laboratory at the Indraprastha Institute of Information Technology, Delhi (IIIT-Delhi), where he is currently working on the GS1 project. He received his Bachelor of Technology degree in Computer Science and Engineering from Jamia Hamdard University, New Delhi, India, in 2023. His research interests include machine learning and large language models, with a focus on developing intelligent and scalable systems for real-world applications.
    
    \hangindent=\dimexpr 1.1in+1em\relax
    \hangafter=-10
    \noindent\llap{\raisebox{\dimexpr\ht\strutbox-\height}[0pt][0pt]{\includegraphics[width=1in,height=1in]{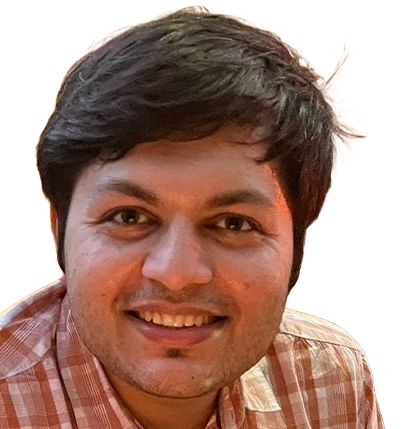}}%
    \hspace{1em}}
    \scriptsize \textbf{Dr. Yaman Kumar Singla} is a Senior Research Scientist at the Adobe Media and Data Science Research (MDSR) laboratory. His research is focused on advancing the state of the art in advertising, marketing, and computational social science. He studies the behavior modality—a framework consisting of Speaker, Content, Channel, Receiver, Behavior (Effect), and temporal factors—to model and optimize human behavior. His work spans behavior simulation, personalization, audience targeting, and speaker and time optimization. More broadly, his research aims to develop models that predict, generate, and explain human behavior while providing insights into human decision-making processes. His publications and ongoing projects on behavior in the wild can be found on his personal webpage \href{https://sites.google.com/view/yaman-kumar/}{here}.
    
    \hangindent=\dimexpr 1.1in+1em\relax
    \hangafter=-10
    \noindent\llap{\raisebox{\dimexpr\ht\strutbox-\height}[0pt][0pt]{\includegraphics[width=1in,height=1in]{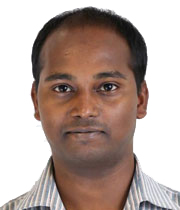}}%
    \hspace{1em}}
    \scriptsize \textbf{Dr. Rajiv Ratn Shah} is an Associate Professor in the Department of Computer Science and Engineering (joint appointment with the Department of Human-centered Design) at IIITD-Delhi. He is the founder of the MIDAS lab at IIITD-Delhi. He received his Ph.D. in Computer Science from the National University of Singapore, Singapore. Dr. Shah has received several awards, including the prestigious Heidelberg Laureate Forum (HLF) and European Research Consortium for Informatics and Mathematics (ERCIM) fellowships. His research interests include multimedia content processing, natural language processing, image processing, multimodal computing, data science, social media computing, and the Internet of Things.

\end{document}